\documentclass[aps, prl, twocolumn, superscriptaddress, preprintnumbers, nofootinbib]{revtex4-1}

\usepackage{amsmath}	
\usepackage{amsthm}		
\usepackage{amssymb}	
\usepackage{eufrak}
\usepackage{datetime}
\usepackage{graphicx}
\usepackage{color}
\usepackage{verbatim}
\usepackage{bm}
\usepackage{color}
\usepackage{bm}
\usepackage{tabularx}
\usepackage[english]{babel}
\usepackage{lipsum}
\usepackage[colorlinks=true, citecolor=midblue, linkcolor=midblue, urlcolor=midblue]{hyperref}

\definecolor{grey}{rgb}{0.4,0.4,0.4}
\definecolor{dullmagenta}{rgb}{0.4,0,0.4}
\definecolor{darkblue}{rgb}{0,0,0.4}
\definecolor{midblue}{rgb}{0,0,0.5}
\definecolor{midred}{rgb}{0.5,0,0}
\definecolor{orange}{rgb}{1,0.5,0}
\definecolor{lightbrown}{rgb}{0.75,0.5,0.25}
\definecolor{tan}{cmyk}{0.14,0.42,0.56,0}
\definecolor{djunglegreen}{cmyk}{0.99,0,0.52,0}
\definecolor{lightgreen}{rgb}{0,1,0}
\definecolor{olivegreen}{cmyk}{0.64,0,0.95,0.40}
\definecolor{midgreen}{rgb}{0.0,0.675,0.0}
\definecolor{darkgreen}{rgb}{0,0.5,0}

\newcommand{\vs}{\vspace}

\renewcommand{\.}{\hspace{0.5mm}}

\newcommand{\la}{\ensuremath{\leftarrow}}

\newcommand{\brm}{\ensuremath{\mathrm{b}}}
\newcommand{\crm}{\ensuremath{\mathrm{c}}}
\newcommand{\drm}{\ensuremath{\mathrm{d}}}
\newcommand{\erm}{\ensuremath{\mathrm{e}}}

\newcommand{\hrm}{\ensuremath{\mathrm{h}}}

\newcommand{\mrm}{\ensuremath{\mathrm{m}}}

\newcommand{\srm}{\ensuremath{\mathrm{s}}}
\newcommand{\trm}{\ensuremath{\mathrm{t}}}

\newcommand{\Mcal}{\ensuremath{\mathcal{M}}}

\newcommand{\Ocal}{\ensuremath{\mathcal{O}}}

\renewcommand{\d}{\ensuremath{\mathrm{d}}}

\newcommand{\MeV}{\ensuremath{\mathrm{MeV}}}

\newcommand{\ie}{i.e.}

\settimeformat{ampmtime}

\usepackage{float}

\newcommand{\be}{\begin{equation}}
\newcommand{\ee}{\end{equation}}
\newcommand{\ba}{\begin{eqnarray}}
\newcommand{\ea}{\end{eqnarray}}

\def\ga{\mathrel{\raise.3ex\hbox{$>$\kern-.75em\lower1ex\hbox{$\sim$}}}}
\def\la{\mathrel{\raise.3ex\hbox{$<$\kern-.75em\lower1ex\hbox{$\sim$}}}}

\def\Msun{M_\odot}

\def\fPBH{f_{\rm PBH}}

\def\fPBH{f_{\rm PBH}}

\newcounter{defcounter}
\newenvironment{AppEquation}{%
\addtocounter{equation}{-1}
\refstepcounter{defcounter}

\begin{equation}}
{\end{equation}$\mspace{-6.5mu}$}

\begin{document}

\title{Cosmic Conundra Explained by Thermal History and Primordial Black Holes}

\author{Bernard Carr}
\email{B.J.Carr@qmul.ac.uk}
\affiliation{School of Physics and Astronomy,
	Queen Mary University of London,
	Mile End Road,
	London E1 4NS,
	UK}
\affiliation{
	Research Center for the Early Universe,
	University of Tokyo,
	Tokyo 113-0033,
	Japan
	}

\author{S{\'e}bastien Clesse}
\email{sebastien.clesse@ulb.ac.be}
\affiliation{
	Service de Physique Th{\'e}orique,
	University of Brussels (ULB),
	Boulevard du Triomphe,
	CP225, 1050, Brussels,
	Belgium}

\author{Juan Garc{\'i}a-Bellido}
\email{juan.garciabellido@uam.es}
\affiliation{
	Instituto de F{\'i}sica Te{\'o}rica UAM-CSIC,
	Universidad Auton{\'o}ma de Madrid,
	Cantoblanco,
	28049 Madrid,
	Spain
	}

\author{Florian K{\"u}hnel}
\email{florian.kuehnel@physik.uni-muenchen.de}
\affiliation{
	Arnold Sommerfeld Center,
	Ludwig-Maximilians-Universit{\"a}t,
	Theresienstra{\ss}e 37,
	80333 M{\"u}nchen,
	Germany}

\date{\formatdate{\day}{\month}{\year}, \currenttime}

\begin{abstract}
A universal mechanism may be responsible for several unresolved cosmic conundra. The sudden drop in the pressure of relativistic matter at $W^{\pm}/Z^{0}$ decoupling, the quark--hadron transition and $e^{+}e^{-}$ annihilation enhances the probability of primordial black hole (PBH) formation in the early Universe. Assuming the amplitude of the primordial curvature fluctuations is approximately scale-invariant, this implies a multi-modal PBH mass spectrum with peaks at $10^{-6}$, 1, 30, and $10^{6}\,\Msun$. This suggests a unified PBH scenario which naturally explains the dark matter and recent microlensing observations, the LIGO/Virgo black hole mergers, the correlations in the cosmic infrared and X-ray backgrounds, and the origin of the supermassive black holes in galactic nuclei at high redshift. A distinctive prediction of our model is that LIGO/Virgo should observe black hole mergers in the mass gaps between 2 and $5\,\Msun$ (where no stellar remnants are expected) and above $65\,\Msun$ (where pair-instability supernovae occur) and low-mass-ratios in between. Therefore the recent detection of events GW190425, GW190814 and GW190521 with these features is striking confirmation of our prediction and may indicate a primordial origin for the black holes. In this case, the exponential sensitivity of the PBH abundance to the equation of state would offer a unique probe of the QCD phase transition. The detection of PBHs would also offer a novel way to probe the existence of new particles or phase transitions with energy between $1\,{\rm MeV}$ and $10^{10}$\,GeV.
\end{abstract}


\maketitle

{\it Introduction{\;---\;}}Primordial black holes (PBHs) in the solar-mass range have attracted a lot of attention since the LIGO/Virgo detection of gravitational waves from coalescing black holes~\cite{Abbott:2016nmj}. The observed merger rate is compatible with what would be expected if PBHs constitute an appreciable fraction, and possibly all, of the cold dark matter (CDM). Moreover, the LIGO/Virgo observations seem to favour mergers with low effective spins, as expected for PBHs but hard to explain for black holes of stellar origin~\cite{Wysocki:2017isg}. An extended PBH mass function with a peak in the range $1$--$10\,\Msun$ could explain the LIGO/Virgo observations. Based on an argument related to gravitational lensing by PBH clusters, we show that the usual dark-matter constraints from the microlensing of stars, supernovae and quasars in this range can be evaded.

Given the revival of interest in PBHs, one must explain why they have the mass and density required for explaining the LIGO/Virgo events, and why these values are comparable to the mass and density of stars. One approach is to choose an inflationary scenario which produces a peak in the power spectrum of curvature fluctuations at the appropriate scale~\cite{GarciaBellido:1996qt}. The required amplitude of the inhomogeneities must be much larger than that observed on cosmological scales but not too large, so this requires fine-tuning of both the scale and amplitude.

An alternative approach is to assume the power spectrum is smooth (\ie~featureless) but that there is a sudden change in the plasma pressure at a particular cosmological epoch, allowing PBHs to form more easily then. Enhanced gravitational collapse occurs because the critical density fluctuation required for PBH formation ($\delta_{\crm}$) decreases when the equation-of-state parameter ($w \equiv p / \rho c^{2}$) is reduced. Since the PBH collapse fraction depends exponentially on $\delta_{\crm}$ for Gaussian fluctuations~\cite{Carr:1975qj}, this can have a strong effect on the fraction of CDM in PBHs. This is particularly important for the Quantum Chromodynamics (QCD) transition at ${\sim} 10^{-5}\,\srm$, lattice-gauge-theory calculations indicating that the sound-speed decreases by around 30\% then~\cite{Schmid:1998mx, Laine:2006cp, Bhattacharya:2014ara, Laine:2015kra, Borsanyi:2016ksw, Byrnes:2018clq, Saikawa:2018rcs}.

PBHs formed at the QCD transition would naturally have the Chandrasekhar mass ($1.4\,\Msun$), this also characterising the mass of main-sequence stars, and a collapse fraction of order the cosmic baryon-to-photon ratio (${\sim} 10^{-9}$) if PBHs provide most of the dark matter~\cite{Carr:2019hud}. The latter feature is naturally explained if PBH formation generates a hot outgoing shower of relativistic particles because electroweak baryogenesis can occur very efficiently there and produce a \textit{local} baryon-to-photon ratio of order unity~\cite{GarciaBellido:2019vlf}.

In this paper we point out an interesting consequence of the above scenario, by extending it beyond the QCD scale. As the background temperature decreases from $100$\,GeV to $1$\,MeV, corresponding to the rest masses of the W and Z bosons, the proton, the pion and the electron, there are \textit{four} periods at which the sound speed exhibits sudden dips. The proton dip is the biggest (${\sim} 30\%$) but the others can also be significant ($5$--10\%) because of the exponential dependence of the gravitational collapse probability on the critical curvature fluctuation. These dips produce distinctive features in the PBH mass function at four mass scales in the range $10^{-6}$--$10^{6}\,\Msun$.

An important feature of this scenario is that it predicts the form of the PBH mass distribution very precisely. We show that for a nearly scale-invariant primordial power spectrum, the expected form not only satisfies all the current astrophysical and cosmological constraints, but also allows the PBHs to explain numerous observational conundra: (1) microlensing events towards the Galactic bulge generated by planet-mass objects with about 1\% of the CDM density~\cite{Niikura:2019kqi}, well above most expectations for free-floating planets; (2) microlensing of quasars~\cite{Mediavilla:2017bok}, including ones that are so misaligned with the lensing galaxy that the probability of lensing by a star is very low; (3) the unexpected high number of microlensing events towards the Galactic bulge by dark objects in the \textit{mass gap} between $2$ and $5\,\Msun$~\cite{Wyrzykowski:2019jyg}, where stellar evolution models fail to form black holes~\cite{Brown:1999ax}; (4) unexplained correlations in the source-subtracted X-ray and cosmic infrared background fluctuations~\cite{2005Natur.438...45K}; (5) the non-observation of ultra-faint dwarf galaxies below the critical radius of dynamical heating by PBHs~\cite{Clesse:2017bsw}; (6) the masses, spins and coalescence rates for the black holes found by LIGO/Virgo~\cite{LIGOScientific:2018mvr}, including two recent events with black holes which are probably in the mass gap; (7) the relationship between the mass of a galaxy and that of its central black hole.
\vs{-0.9mm}

{\it Thermal History of the Universe{\;---\;}}Reheating at the end of inflation fills the Universe with radiation. In the absence of extensions beyond the Standard Model (SM) of particle physics (eg.~with right-handed neutrinos), the Universe remains dominated by relativistic particles with an energy density decreasing as the fourth power of the temperature as the Universe expands. The number of relativistic degrees of freedom remains constant ($g_{*} = 106.75$) until around $200$\,GeV, when the temperature of the Universe falls to the mass thresholds of SM particles.

\begin{figure}
	\centering
	\vs{-3mm}
	\includegraphics[width = 0.48\textwidth]{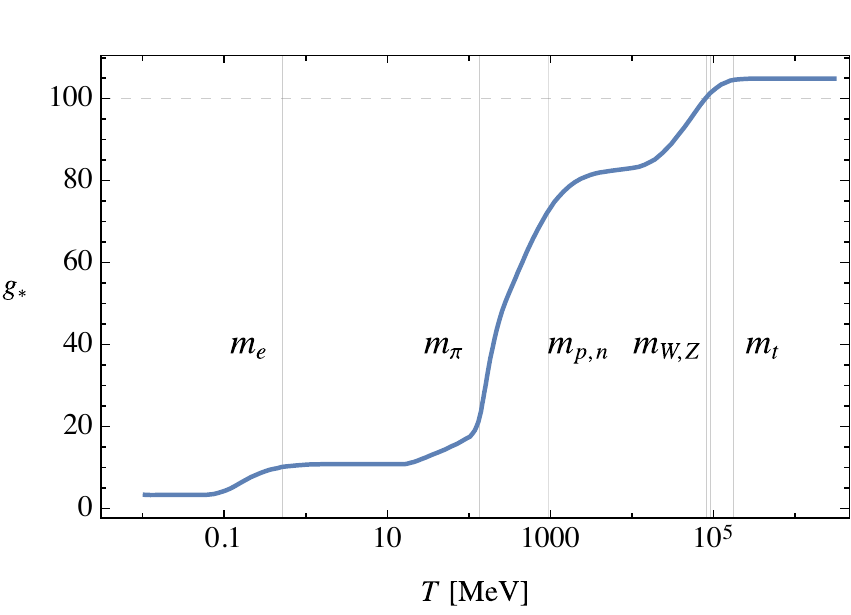}
	\includegraphics[width = 0.48\textwidth]{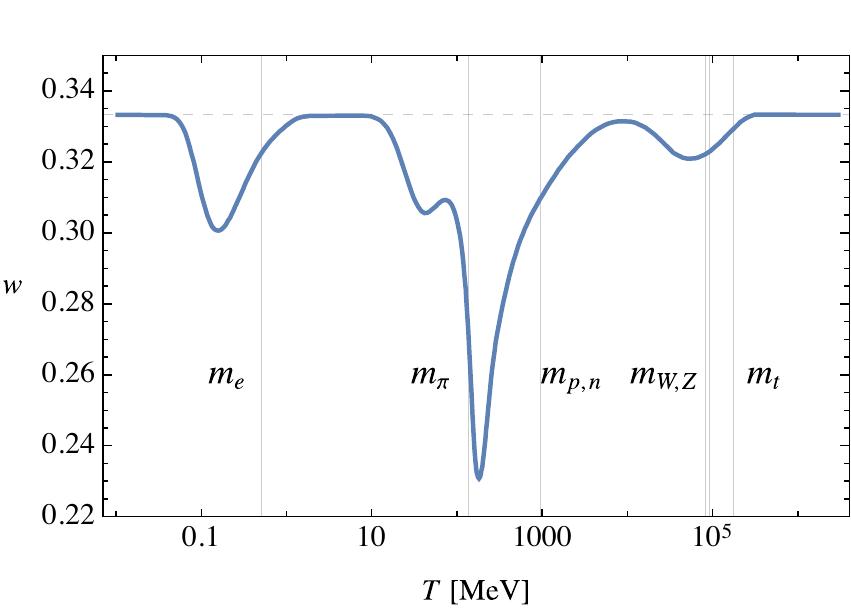}
	\caption{Relativistic degrees of freedom $g_{*}$ ({\it upper panel}) 
			and equation-of-state parameter $w$ ({\it lower panel}), 
			both as a function of temperature $T$ (in MeV).
			The grey vertical lines correspond to the masses of the electron, 
			pion, proton/neutron, $W,\.Z$ bosons and top quark, respectively.
			The grey dashed horizontal lines indicate values of $g_{*} = 100$ 
			and $w = 1 / 3$, respectively.}
	\label{fig:g-and-w-of-T}
\end{figure}

As shown in Fig.~\ref{fig:g-and-w-of-T} (upper panel), the first particle to become non-relativistic is the top quark at $T \simeq m_{\trm} = 172\,$GeV, followed by the Higgs boson at $125\,$GeV, and the $Z$ and $W$ bosons at $92$ and $81\,$GeV, respectively. These particles become non-relativistic at nearly the same time and this induces a significant drop in the number of relativistic degrees of freedom down to $g_{*} = 86.75$. There are further changes at the $b$ and $c$ quark and $\tau$-lepton thresholds but these are too small to appear in Fig.~\ref{fig:g-and-w-of-T}. Thereafter $g_{*}$ remains approximately constant until the QCD transition at around $200\,$MeV, when protons and neutrons condense out of the free light quarks and gluons. The number of relativistic degrees of freedom then falls abruptly to $g_{*} = 17.25$. A little later the pions become non-relativistic and then the muons, giving $g_{*} = 10.75$. Thereafter $g_{*}$ remains constant until $e^{+}e^{-}$ annihilation and neutrino decoupling at around $1\,$MeV, when it drops to $g_{*} = 3.36$.

Whenever the number of relativistic degrees of freedom suddenly drops, it changes the effective equation of state parameter $w$. As shown in Fig.~\ref{fig:g-and-w-of-T} (lower panel), there are thus four periods in the thermal history of the Universe when $w$ decreases. After each of these, $w$ resumes its relativistic value of $1/3$ but each sudden drop modifies the probability of gravitational collapse of any large curvature fluctuations present at that time. We will see below how these changes in $w$ result in the production of PBHs with different masses and dark-matter fractions.

The above discussion is subject to some uncertainties. Firstly, a proper description of phase transitions requires highly perturbative numerical methods which are not yet fully understood; deviations in $w$ of a few percent are possible~\cite{Laine:2015kra, Borsanyi:2016ksw, Saikawa:2018rcs, Vovchenko:2020crk}. Secondly, before the onset of neutrino oscillations, significant non-zero lepton flavour asymmetries may be present~\cite{Oldengott:2017tzj}, leading to significant changes in the background plasma pressure~\cite{Middeldorf-Wygas:2020glx, Vovchenko:2020crk, Bodeker:2020stj}. In particular, the Universe can pass through a pion condensate phase for large lepton flavour asymmetry~\cite{Vovchenko:2020crk}. This softens the equation of state more than indicated in Fig.~\ref{fig:g-and-w-of-T} and the QCD transition may even become 1st order~\cite{Middeldorf-Wygas:2020glx}. However, in this work we assume zero lepton flavour asymmetry and base our results on Ref.~\cite{Borsanyi:2016ksw}.

{\it Primordial Black Hole Formation{\,---\,}}There are a plethora of mechanisms for PBH formation. All of them require the generation of large overdensities, specified by the density contrast, $\delta \equiv \delta \rho / \rho$, usually assumed to be of inflationary origin. When overdensities re-enter the Hubble horizon, they collapse if they are larger than some threshold $\delta_{\crm}$, which generally depends on the equation of state and density profile. However, there are other (non-inflationary) scenarios for PBH formation, where the inhomogeneities arise from first-order phase transitions, bubble collisions, and the collapse of cosmic strings, necklaces, domain walls or non-standard vacua. Full references can be found in Ref.~\cite{Carr:2020xqk}.

The threshold $\delta_{\crm}$ is a function of the equation-of-state parameter $w( T )$, which is shown in Fig.~\ref{fig:g-and-w-of-T}, so the thermal history of the Universe can induce pronounced features in the PBH mass function even for a uniform power spectrum. This is because, if the PBHs form from Gaussian inhomogeneities with root-mean-square amplitude $\delta_{\rm rms}$, then the fraction of horizon patches undergoing collapse to PBHs when
the temperature of the Universe is $T$ should be~\cite{Carr:1975qj}
\be
	\beta( M )
		\approx
				{\rm erfc}\!
				\left[
					\frac{\delta_{\crm}\big( w[ T( M ) ] \big)}
					{ \sqrt{2} \, \delta_{\rm rms}( M )}
				\right]
				,
				\label{eq:beta(T)}
\ee
where `erfc' is the complementary error function and the temperature is related to the PBH mass by 
\be
	T
		\approx
				200\,
				\sqrt{\Msun / M\,}\;\MeV
				\, .
\ee
This shows that $\beta( M )$ is exponentially sensitive to $w( M )$. Throughout this work, we use the numerical results for $\delta_{\crm}$ from Ref.~\cite{Musco:2012au}. We stress that there are theoretical uncertainties in the value of $\delta_{\crm}$, associated (for example) with the density profile in the collapsing region~\cite{Musco:2020jjb, Young:2014ana, Yoo:2018kvb, Musco:2018rwt, Kuhnel:2015vtw}. However, any change in its value can be counterbalanced by an adjustment in $\delta_{\rm rms}$, such that the current PBH dark matter fraction is preserved.

We need an expression for the present fraction of the CDM in PBHs with mass $M$. However, this requires clarification since the density of PBHs with a precise mass is not defined for a continuous mass function. If the number density of PBHs in the mass range $(M,\,M + \d M)$ is $\d n$, then one can define the mass density and dark matter fraction of PBHs with mass `around' $M$ (\ie~in the mass range $M$ to $2M$) by
\be
	\rho_{\rm PBH}( M )
		\equiv
				M^{2}\.
				\frac{ \d n }
				{ \d M }
				\, ,
	\quad
	f_{\rm PBH}
		=
				\frac{ \rho_{\rm PBH}( M ) }
				{ \rho_{\rm CDM} }
				\, ,
\ee
where $\rho_{\rm CDM}$ is the CDM density. Integrating this over $M$ gives the {\it total} dark matter fraction $f^{\rm tot}_{\rm PBH}$. The present CDM fraction for PBHs with mass around $M$ is then
\be
	\fPBH( M ) 
				\approx 2.4\,\beta( M )\.
				\sqrt{
					\frac{ M_{\rm eq} }
					{ M }
				\,}
				\, ,
				\label{eq:fPBH}
\ee
where $M_{\rm eq} = 2.8 \times 10^{17}\,\Msun$ is the horizon mass at matter-radiation equality. The numerical factor is $2\.( 1 + \Omega_{\brm} / \Omega_{\rm CDM} )$, with $\Omega_{\rm CDM} = 0.245$ and $\Omega_{\brm} = 0.0456$ being the CDM and baryon density parameters \cite{Aghanim:2018eyx}.

There are many inflationary models and these predict a variety of shapes for $\delta_{\rm rms}( M )$. Some of them, including single-field models like Higgs inflation \cite{Ezquiaga:2017fvi} or two-field models like hybrid inflation \cite{Clesse:2015wea}, produce an extended plateau or dome-like feature in the power spectrum. Instead of focussing on any specific scenario, we will assume a spectrum of the form
\be
	\delta_{\rm rms}( M )
		=
				\begin{cases}
					\tilde{A}
					\left(
						\frac{ M }{ \Msun }
					\right)^{\!(1 - \tilde{n}_{\srm}) / 4}
					&\; \text{(PBH scales)} \\[2mm]
					A
					\left(
						\frac{ M }{ \Msun }
					\right)^{\!(1 - n_{\srm}) / 4}
					&\; \text{(CMB scales).} 
				\end{cases}
				\label{eq:delta-power-law}
\ee
Here the spectral index $n_{\srm}$ and amplitude $A$ are taken to have their CMB values~\cite{Aghanim:2018eyx}, $n_{\srm} = 0.97$ and $A = 4 \times 10^{-5}$, respectively, while the corresponding small-scale quantities, $\tilde{n}_{\srm}$ and $\tilde{A}$, are treated as free phenomenological parameters. Indeed, Eq.~\eqref{eq:delta-power-law} can represent any spectrum with an additional broad peak or small-scale enhancement, such as might be generically produced by a second phase of slow-roll inflation. In order to get an integrated PBH abundance of $\fPBH^{\rm tot} = 1$, the small-scale amplitude has to be $\tilde{A} = 0.1487$ if $\tilde{n}_{\srm} = n_{\srm}$. As discussed in Appendix A1, non-Gaussian and non-linear effects~\cite{Kalaja:2019uju} can impact the overall PBH abundance, but one can rescale $\tilde{A}$ to give $\fPBH^{\rm tot} = 1$ without significantly affecting the mass function. 

\begin{figure}
	\vs{-3mm}
	\centering
	\includegraphics[width = 0.48\textwidth]{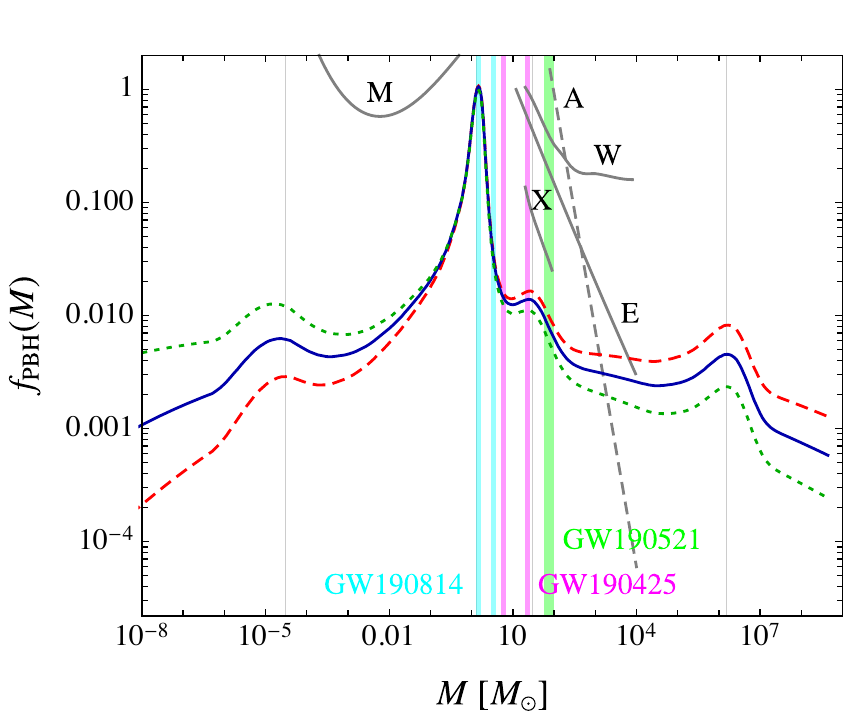}
	\vs{-2mm}
	\caption{The mass spectrum of PBHs
			[see Eq.~\eqref{eq:delta-power-law}] with spectral index 
			$\tilde{n}_{\srm} = 0.955$ (red, dashed),
			$0.960$ (blue, solid),
			$0.965$ (green, dotted).
			The grey vertical lines corresponds to the EW and QCD phase transitions 
			and $e^{+}e^{-}$ annihilation. 
			The vertical coloured lines indicate the masses of the
			three recent LIGO-Virgo events.
			Also shown (grey curves) are constraints from 
			microlensing (M) with the assumptions of Table~\ref{tab:fDM} (fourth line), 
			ultra-faint dwarf galaxies and Eridanus II (E) \cite{Li:2016utv}, 
			X-ray/radio counts (X) \cite{Gaggero:2016dpq}, and
			halo wide binaries (W) \cite{Quinn:2009zg}. 
			The accretion constraint (A) \cite{Ali-Haimoud:2016mbv} is shown dashed because 
			it relies on uncertain astrophysical assumptions (see Table~\ref{tab:fDM}).
			}
	\label{fig:fPBH}
\end{figure}

The ratio of the PBH mass and the horizon mass at re-entry is denoted by $\gamma$ and we assume $\gamma = 0.7$ as a benchmark value, following Ref.~\cite{GarciaBellido:2019vlf}. The resulting mass function is represented in Fig.~\ref{fig:fPBH}. It exhibits a dominant peak at $M \simeq 2\,\Msun$ and three additional bumps at $10^{-5}\,\Msun$, $30\,\Msun$ and $10^{6}\,\Msun$, corresponding to transitions in the number of relativistic degrees of freedom predicted by the known thermal history of the Universe.\footnote{We also indicate the values of $M$ associated with the three recent LIGO/Virgo events, although this data only became available after our prediction.} Jedamizk~\cite{Jedamzik:1996mr} first drew attention to the dips at the QCD and electron-annihilation epochs. Byrnes {\it et al.}~\cite{Byrnes:2018clq} have derived the PBH mass function associated with the QCD transition but this omits the smallest and largest mass bumps. We note that the prominence of the latter depends on the tilt of the power spectrum and our choice of tilt makes it more significant. Byrnes {\it et al.} also incorporate a more detailed study of the effects of criticality on the low-mass tail of the PBH mass function. Although the time dependence of the equation of state of the thermal plasma is crucial in determining the PBH mass function, the tail effect is not large; it just lowers and slightly broadens the main peak at around $1\,\Msun$.
\vs{0.5mm}

{\it Constraints{\;---\;}}
In this section, we discuss whether the PBH mass functions shown in Fig.~\ref{fig:fPBH}, all of which assume $\fPBH^{\rm tot} = 1$, are compatible with the numerous observational constraints on $\fPBH( M )$. There is an underproduction of light PBHs for $\tilde{n}_{\srm} \leq 0.955$ and of heavy ones for $\tilde{n}_{\srm} \geq 0.965$ but we claim the mass distribution for $\tilde{n}_{\srm} \simeq 0.96$ can provide 100\% of the dark matter without violating any current reliable constraints, despite some claims to the contrary.

In order of increasing mass, the PBH constraints come from the extragalactic $\gamma$-ray background, microlensing surveys, dynamical effects (such as the heating of ultra-faint dwarf galaxies and their stellar clusters and the disruption of wide binaries), ratio and X-ray point source counts, and CMB anisotropies generated by PBH accretion. The limits are summarised in Ref.~\cite{Carr:2016drx} and numerous other papers. Most of these constraints assume a monochromatic PBH mass function (\ie~one with width $\Delta M \sim M$). In the present scenario we predict an extended mass function and cannot simply compare this with the monochromatic constraints. In order to assess the situation, we adopt the approach advocated in Ref.~\cite{2017PhRvD..96b3514C}. Assuming that the mass distribution scales linearly with $\fPBH^{\rm tot}$, each probe $p$ sets an upper limit
\be
	\fPBH^{\rm max}
		=
				\left(
					\int\!\drm\mspace{-1mu}\ln M\;
					\frac{\fPBH( M ) }
					{ f_{p}^{\rm mon}( M )}
				\right)^{\!-1}
				\label{Eq:fmax}
				,
\ee
where $f_{p}^{\rm mon}( M )$ is the limit from probe $p$ for a monochromatic function of mass $M$. We have calculated the value of $\fPBH^{\rm max}$ associated with each probe for $\tilde{n}_{\srm} = 0.96$ but different astrophysical assumptions. These are shown in Table~\ref{tab:fDM} and discussed in more detail in the appendices material.

It is sometimes argued that the EROS/MACHO microlensing limits exclude solar-mass PBHs with $\fPBH^{\rm tot} = 1$ but this is based on various contentious assumptions (monochromatic mass function, no clustering, isothermal halo profile). In Appendix A2, we show that these limits are evaded in our scenario. This is because the primordial power spectrum is enhanced on small scales, so the corresponding inhomogeneities virialise much more quickly than in the standard scenario, with the PBHs forming compact clusters. The CMB limits of Refs.~\cite{Ali-Haimoud:2016mbv, Poulin:2017bwe} are still in tension with our model but only for $M \ga 10^{3}\,\Msun$ and the steady-state accretion assumption breaks down for such large masses. The highest mass peak in Fig.~\ref{fig:fPBH} also conflicts with the CMB $\mu$-distortion limit~\cite{Byrnes:2018txb} (not shown in Fig.~\ref{fig:fPBH}) but one can avoid this by invoking large non-Gaussianity~\cite{Nakama:2017xvq}. This changes the tails of the fluctuation distribution and typically increases the probability of collapse without changing the shape of the mass function~\cite{Ezquiaga:2018gbw} if the distribution above the critical threshold is exponentially suppressed. It is even possible that PBHs formed without an enhancement of the primordial power spectrum due to quantum diffusion~\cite{Ezquiaga:2019ftu}. One can also evade these limits if the transition in the primordial power spectrum indicated by Eq.~\eqref{eq:delta-power-law} occurs at sufficiently small scales but one cannot then
explain the relation between the galactic halo and central black hole masses.

\begin{table*}
	\begin{tabular}{|l|c|l|l|}
	\hline 
	Probe
		& $f_{\rm PBH}^{\rm tot}$
		& Assumptions
		& References\vphantom{$1^{^{1}}$}\\[0.5mm]
	\hline 
	\hline 
	\vphantom{$1^{^{1}}$}
		& $0.08$
		& isothermal dark matter halo profile, no clustering
		& EROS \cite{Tisserand:2006zx}, OGLE \cite{Wyrzykowski:2009ep}\\
	{\bf Microlensing}	
		& $0.39$
		& conservative isothermal profile, no clustering 
		& Green \cite{Green:2016xgy}\\
	{\bf (LMC/SMC)}
		& $0.16$
		& realistic profile, no clustering
		& Calcino {\it et al.} \cite{Calcino:2018mwh}\\
		& $3.9$ 
		& Green \cite{Green:2016xgy} model plus $90\%$ of PBHs in clusters $> 10^{3}\,\Msun$
		& appendices \\[1.0mm]
	{\bf Eridanus II }
		& $2.3$
		& IMBH of $M = 2000\,\Msun$ at centre
		& Li {\it et al.} \cite{Li:2016utv}, Brandt \cite{Brandt:2016aco}\\[0.8mm]
	{\bf X-ray/radio PS in GC}
		& 4.5
		& accretion parameter $\lambda = 0.02$, 
			Maxwellian velocity distribution 
		& Gaggero {\it et al.} \cite{Gaggero:2016dpq}\\[1.0mm]
	{\bf Halo wide binaries}
		& $19$
		& fixed relative velocity, constant dark matter profile
		& Quinn {\it et al.} \cite{Quinn:2009zg}\\[1.0mm]
	{\bf Planck}
		& $0.17$
		& conservative case, steady-state accretion, no clustering
		& Ali-Ha{\"i}moud {\it et al.} \cite{Ali-Haimoud:2016mbv}\\[1.0mm]
	\hline
	\end{tabular}
	\caption{Upper limits (at $95\%$\,C.L.) on the integrated PBH fraction 
			$\fPBH^{\rm tot}$ from the various probes discussed in the text, 
			calculated for a PBH mass function with $\tilde{n}_{\srm} = 0.96$. 
			The third column lists the main assumptions and uncertainties 
			underlying these limits. The PBHs can account for all the dark matter 
			if one uses a realistic estimate of the EROS microlensing constraints 
			(fourth line). Uncertainties in CMB limits for this mass function are 
			discussed in the text.
			\vs{-3mm}}
	\label{tab:fDM}
\end{table*}

{\it Observational Conundra{\;---\;}}Besides passing the current observational constraints on the form of the CDM, the PBH mass function with $\tilde{n}_{\srm} \simeq 0.96$ predicted from the known thermal history of the Universe provides a unified explanation for several other puzzling conundra. We discuss these in order of increasing PBH mass. The status of some of the conundra is still unclear but we include all of them to convey the breadth of predictions.

{\it 1.~Planetary-Mass Microlenses.} Recently Niikura {\it et al.}~have reported two interesting microlensing results. The first \cite{Niikura:2017zjd} comes from observations of M31 using the Subaru telescope, which include one possible detection and place strong constraints on PBHs in the mass range $10^{-10}$ to $10^{-6}\,\Msun$. The constraints are roughly compatible with our model and even the single candidate could be. The second \cite{Niikura:2019kqi} uses data from the five-year OGLE survey of 2622 microlensing events in the Galactic bulge \cite{2017Natur.548..183M} and has revealed six ultra-short ones attributable to planetary-mass objects between $10^{-6}$ and $10^{-4}\,\Msun$. These would contribute about $1\%$ of the CDM, which is more than expected for free-floating planets \cite{2019A&A...624A.120V}. This corresponds to the first bump in our predicted PBH mass function and the abundance, when integrated over the mass range probed by OGLE, coincides with our best-fit model with $\tilde{n}_{\srm} \simeq 0.96$.

{\it 2.~Quasar Microlensing.} Hawkins has claimed for many years that quasar microlensing data suggest the existence of PBH dark matter \cite{1993Natur.366..242H}. He originally argued for Jupiter-mass PBHs but later increased the mass estimate to $0.4\,\Msun$ \cite{Hawkins:2006xj}. More recently, the detection of $24$ microlensed quasars \cite{Mediavilla:2017bok} suggests that up to $25\%$ of galactic halos could be in PBHs with mass between $0.05$ and $0.45\,\Msun$ (somewhat below our main peak). These events could also be explained by intervening stars, but in several cases the stellar region of the lensing galaxy is not aligned with the quasar, which suggests a population of subsolar halo objects with $\fPBH > 0.01$. Indeed, Hawkins has argued that the most plausible microlensers are PBHs, either in galactic halos or distributed along the lines of sight to the quasars \cite{Hawkins:2020zie}. For a PBH mass function with $\tilde{n}_{\srm} = 0.96$, one expects $\fPBH \simeq 0.07$ in this mass range. In principle, Ref.~\cite{Mediavilla:2017bok} excludes all the dark matter being in the main peak at $2\,\Msun$ but that conclusion can be circumvented if the PBHs are in clusters and we argue in Appendix A2 that only $10\%$ of them should be uniformly distributed.

{\it 3.~OGLE/GAIA Excess of Dark Lenses in the Galactic Bulge.} OGLE has detected around $60$ long-duration microlensing events, of which around $20$ have GAIA parallax measurements which break the mass-distance degeneracy and imply that they are probably black holes \cite{Wyrzykowski:2019jyg}. The event distribution from the posterior likelihood of their masses peaks between $0.8$ and $5\,\Msun$, which overlaps the gap from $2$ to $5\,\Msun$ in which black holes are not expected to form as the endpoint of stellar evolution \cite{Brown:1999ax}. Although most of the dark matter is not in this mass gap, this is consistent with the main peak in the PBH mass distribution if $0.6 \lesssim \gamma \lesssim 1$.

{\it 4.~Cosmic Infrared/X-ray Backgrounds.} As shown by Kashlinsky \cite{2005Natur.438...45K, Kashlinsky:2016sdv}, 
the spatial coherence of the X-ray and infrared source-subtracted backgrounds implies that black holes are required. Although these need not be primordial, the level of the infrared background suggests an overabundance of high-redshift halos and this could be explained if a significant fraction of the CDM comprises PBHs larger than a few solar masses, the Poisson fluctuations in their number density then growing all the way from matter-radiation equality. In these halos, a few stars form and emit infrared radiation, while PBHs emit X-rays due to accretion. It is challenging to find other scenarios that naturally produce such features. The required mass cannot be specified precisely. Ref.~\cite{Kashlinsky:2016sdv} focuses on $30\,\Msun$ PBHs and the Poissonian power is smaller by an order of magnitude for $3\,\Msun$ PBHs. However, this only reduces the mass of high-$z$ halos by a factor of a few, so the IR background can still be explained with our mass distribution.

{\it 5.~Ultra-Faint Dwarf Galaxies (UFDGs).} For the PBH mass distribution shown in Fig.~\ref{fig:fPBH}, the critical radius below which CDM-dominated UFDGs would be dynamically unstable is $r_{\crm} \sim 10$--$20$~parsecs (depending on the mass of a possible central black hole). The non-detection of galaxies smaller than this critical radius, despite their magnitude being above the detection limit, suggests compact halo objects in the solar-mass range. Moreover, rapid accretion in the densest PBH halos could explain the extreme UFDG mass-to-light ratios observed \cite{Clesse:2017bsw}. Recent $N$-body simulations \cite{Boldrini:2019isx} confirm that this mechanism works for PBHs of $25$--$100\,\Msun$ providing they provide at least $1$\% of the dark matter.

\begin{figure}
	\vspace{-2mm}
	\centering 
	\includegraphics[width = 0.5\textwidth]{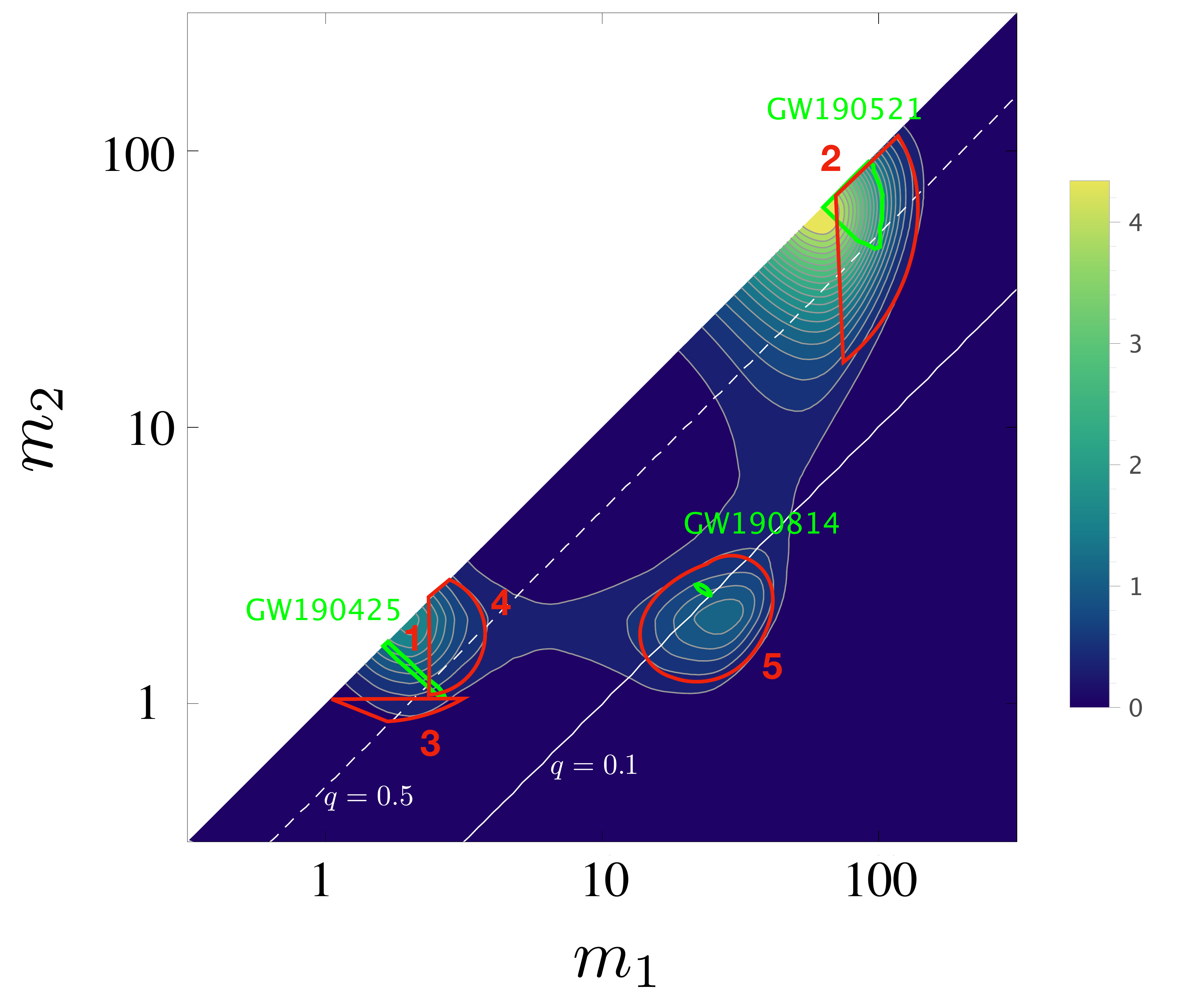}
	\vs{-5mm}
	\caption{
			Expected probability distribution of PBH merger detections with masses 
			$m_{1}$ and $m_{2}$ (in units of solar mass)
			by LIGO/Virgo, assuming a PBH mass function with $\tilde{n}_{\srm} = 0.96$, 
			based on the LIGO spectral noise density for the O2 run and 
			the method described in Appendix A3.
			The solid and dashed white lines correspond to mass ratios 
			$q = m_{2} / m_{1}$ of $0.1$ and $0.5$, respectively.
			The coloured sidebar gives the relative probability. 
			The peak of our distribution at (1)
				would be taken to be neutron-star mergers 
				without electromagnetic counterparts.
			Stellar black-hole mergers are not expected within the red bounded regions, 
			which are:
			(2)~events above $60\,\Msun$;
			(3)~mergers with a subsolar light component ($m_{2}$)
				and a heavy component 
			($m_{1}$) at the peak of our distribution;
			(4)~mergers with $m_{1}$ in the mass gap;
			(5)~a sub-dominant population of mergers with low mass ratios.
			The three recent LIGO/Virgo detections, 
			which postdate the rest of the figure, 
			are shown in green and lie in regions 2, 4 and 5.
			\vs{-4mm}}
	\label{fig:Rate-LIGO-events}
\end{figure}

{\it 6.~Mass, Spin and Merger Rates for LIGO/Virgo Black Holes.} Most of the observed coalesced black holes have effective spins compatible with zero~\cite{LIGOScientific:2018jsj}. Although the statistical significance
of this result is still low, this goes against a stellar binary origin but is a prediction of the PBH scenario. We assume that most of the binaries form at late times rather than primordially and the reasons for this
(somewhat controversial assumption) are given in Appendix A3. In this case, the expected rate of PBH mergers is comparable to that observed if PBHs account for a significant fraction of the CDM. With our mass distribution, PBHs in the range $10$--$100\,\Msun$ have $\fPBH( M ) \sim 0.03$ even if $\fPBH^{\rm tot} = 1$. Using a method discussed in Appendix A3, we have computed the likelihood distribution of merger events with PBHs of masses $m_{1}$ and $m_{2}$ for $\tilde{n}_{\srm} = 0.96$. The results are shown in Fig.~\ref{fig:Rate-LIGO-events}, which shows that the LIGO/Virgo events are in the most likely region. This region might shift to slightly lower masses if one takes into account the difference between the source frame and detector frame mass for the most distant black hole mergers. With the expected number of O3 events, LIGO/Virgo should be able to detect mergers of PBHs larger than $65\,\Msun$ or with a low mass ratio, $q \equiv m_{2} / m_{1} \lesssim 0.1$, these being distinctive predictions of our scenario. The expected PBH merger rate in the solar-mass range, after normalising to the observed rate of $50\,{\rm yr}^{-1}\,{\rm Gpc}^{-3}$ in the larger mass range, is $\tau \approx 10^{3}\,{\rm yr}^{-1}\,{\rm Gpc}^{-3}$ for PBHs between $1$ and $5\,\Msun$, which is below the rate inferred for neutron-star mergers but within range of the current LIGO/Virgo runs. Our scenario could therefore be probed by searching for BH mergers in the $2$--$5\,\Msun$ mass gap or below the Chandrasekhar mass. These could be distinguished from neutron-star mergers using the maximum chirp frequency or non-detection of electromagnetic counterparts. The LIGO/Virgo collaboration has announced the probable detection of two BH mergers (GW190425 and GW190814) with one component in the mass gap~\cite{Abbott:2020uma, Abbott:2020khf}. These populate regions 4 or 5 of Fig.~\ref{fig:Rate-LIGO-events} and are consistent with our model. The first event~\cite{Abbott:2020uma} could be a merger of PBHs at the ``proton'' peak, given that no electromagnetic counterpart was observed and one component has a mass above that expected for a neutron star. The second event corresponds to the ``pion'' plateau and a recent paper by Jedamzik~\cite{Jedamzik:2020omx} also supports this conclusion. Subsequently, LIGO/Virgo has announced the detection of a black hole merger with at least one component in the pair-instability mass gap (GW190521)~\cite{Abbott:2020tfl, Abbott:2020mjq}, corresponding to region 2 of Fig.~\ref{fig:Rate-LIGO-events}.

\begin{figure}
	\centering
	\includegraphics[width = 0.43\textwidth]{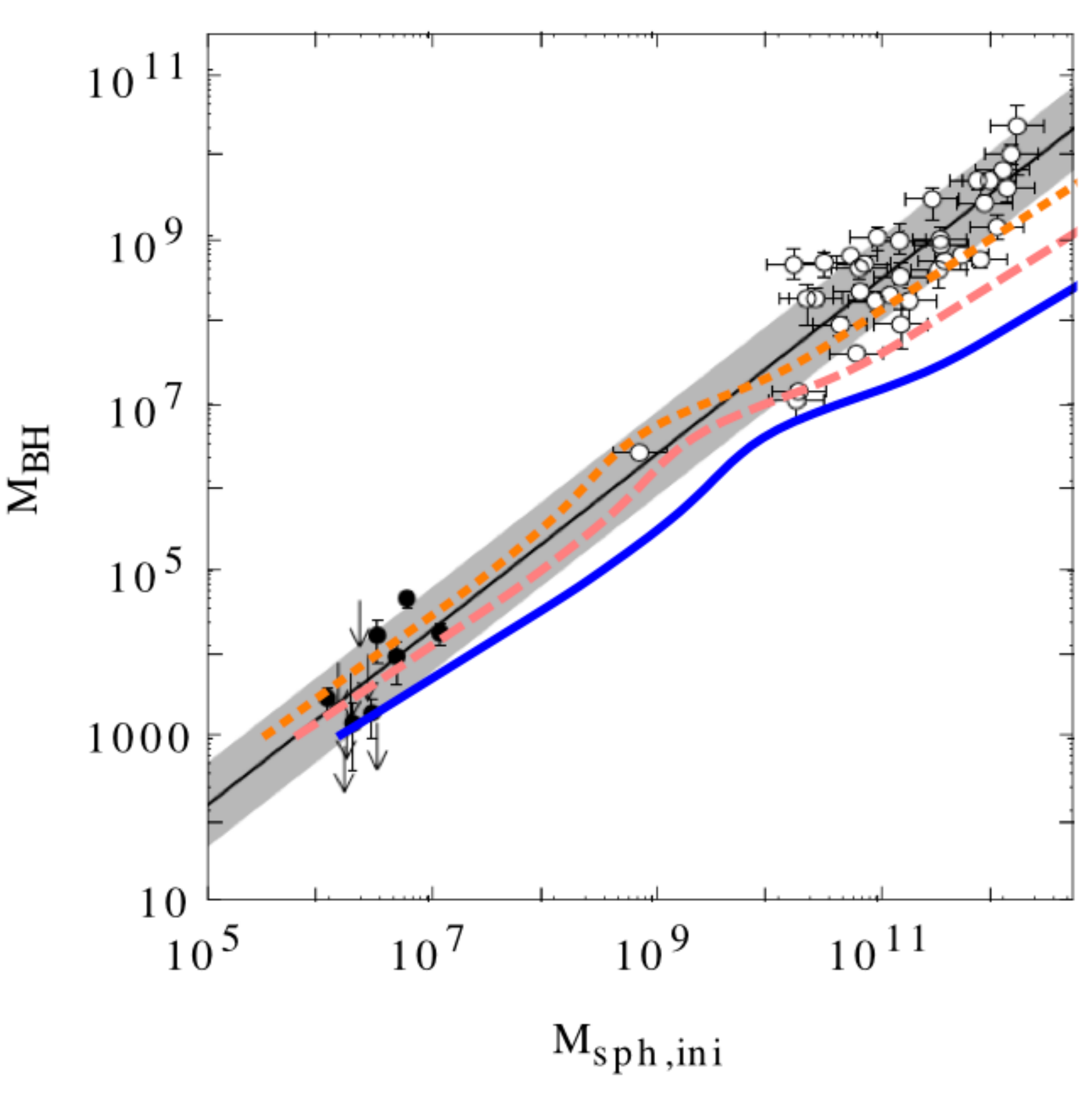}
	\caption{Expected relation between PBH mass and 
			initial host spheroidal mass 
			(in units of solar mass) 
			for 
			$\tilde{n}_{\srm} = 0.95$ (dotted, orange), 
			$\tilde{n}_{\srm} = 0.96$ (dashed, pink) and 
			$\tilde{n}_{\srm} = 0.97$ (solid, blue). 
			Figure and data points adapted from Ref.~\cite{Kruijssen:2013cna}.}
	\label{fig:SMBHratio}
\end{figure}

{\it 7.~IMBHs and SMBHs.} Given our mass distribution, we have calculated the number of intermediate-mass and supermassive PBHs for each $10^{12}\,\Msun$ halo. Interestingly, we obtain about one $10^{8}\,\Msun$ PBH per halo and $10$ times as many smaller ones, possibly seeding the formation of a comparable number of dwarf satellites and faint CDM halos. Assuming a standard Press-Schechter halo mass function \cite{Press:1973iz},
\vs{-2mm}
\be
	\frac{ \drm\.n_{\hrm} }{ \drm \ln M_{\hrm} }
		\approx
				\frac{\rho_{\mrm} }
				{\sqrt{\pi\,}\.
				M_{\hrm}}\,
				\erm^{- M_{\hrm} / M_{*}}
				\; ,
\ee
where $\rho_{\mrm}$ is the mean cosmological matter density (both dark and baryonic) and $M_{*} \approx 10^{14}\,\Msun$ is the cut-off halo mass. For a given $M_{\hrm}$, one can thus identify the corresponding PBH mass that has the same number density, 
\be
	\frac{ \drm\.n_{\rm PBH} }
	{ \drm \ln M }
		\approx
				\frac{ \rho_{\mrm}\.\fPBH }
				{ M }
				\; .
\ee
This gives a relation $M_{\hrm} \approx M_{\rm PBH} / \fPBH$, corresponding to roughly one IMBH/SMBH per halo of mass $10^{3}\,M_{\rm PBH}$ for our distribution, which is in agreement with observations. Furthermore, a mass distribution with $\tilde{n}_{\srm} \approx 0.96$ reproduces the observed relation between the central black hole mass and halo mass \cite{Kruijssen:2013cna}, as shown in Fig.~\ref{fig:SMBHratio}, but only if $\fPBH^{\rm tot} \simeq 1$. A lower (larger) value of the spectral index would imply too many (few) IMBH/SMBHs. Accretion should increase the mass of heavier SMBHs somewhat, and this would make the case $\tilde{n}_{\srm} \approx 0.96$ in closer agreement with observations. Although the initial peak is at $10^{6}\,\Msun$, PBHs of this mass would inevitably grow as a result of accretion.
\vs{1.0mm}

{\it Conclusions{\;---\;}}Various cosmic conundra are naturally explained by the PBH mass function expected from the known thermal history of the Universe if $\fPBH^{\rm tot} = 1$, \ie~if PBHs constitute all of the dark matter. The current LIGO/Virgo run should measure the mass function of coalescing black holes rather precisely and, remarkably, two recent events coincide with the ``proton" peak at around $1\,\Msun$, while a third corresponds to the ``pion" plateau at around $50\,\Msun$. This is indicated in Fig.~\ref{fig:fPBH} and was a \textit{prediction} of our model. PBHs from the ``W/Z'' bump are too small to be seen by LIGO/Virgo but they would be detectable by microlensing effects and may indeed have already been found in OGLE data. PBHs from the ``electron'' bump are too large to be seen by LIGO/Virgo but may be detected by their dynamical effects and may explain the relation between the masses of IMBHs in dwarf spheroidals or SMBHs in galactic nuclei and the masses of the host halos.

It is intriguing that extrapolating the physics of elementary particles back to the early Universe not only resolves the mystery of the dark matter but also addresses so many other cosmic conundra. On the other hand, if firm evidence for PBHs were found, a broad variety of astronomical observations~\cite{Ali-Haimoud:2019khd} could search for extra features in their mass function. This would probe the existence of any new particles thermally coupled to the primordial plasma, independently of their coupling to SM particles, for masses from $1$\,MeV to $10^{10}$\,GeV (above which PBHs should have evaporated), far beyond the energies accessible by any future particle accelerator. 

The exponential sensitivity of the PBH abundance to the equation of state also means that PBHs can be used to probe the characteristics of the cosmic phase transitions at which they form. This is particularly relevant to the detection of gravitational waves from PBHs. For example, we have seen that the LIGO/Virgo results may probe the QCD transition and the presence of lepton flavour asymmetries associated with a pion condensation phase. It is also possible that NANOGrav may have detected a stochastic gravitational wave background and several groups have argued that this could be a 2nd-order background associated with PBH formation~\cite{Kohri:2020qqd, DeLuca:2020agl, Vaskonen:2020lbd, Domenech:2020ers}. If this interpretation of the data were confirmed, this would qualify as another important observational conundrum but we not discuss it further here.

\acknowledgements
{\it Acknowledgements{\;---\;}}The authors thank Y.~Ali-Ha{\"i}moud, C.~Byrnes, A.~Green, M.~Hawkins, K.~Jedamzik, A.~Kashlinsky, P.~Mroz, J.~Rich, C.~Ringeval, D.~Schwarz and Di Wen for stimulating discussions. B.C.~thanks the Research Center for the Early Universe (RESCEU) at University of Tokyo for hospitality received during this work. J.G.-B.~acknowledges support from the Research Project FPA2015-68048-03-3P (MINECO-FEDER) and the Centro de Excelencia Severo Ochoa Program SEV-2016-0597. The work of S.C.~is supported by the Belgian Fund for Research F.R.S.-FNRS and the FRAG team of UCLouvain. F.K.~acknowledges support from the Swedish Research Council (Vetenskapsr{\r{a}}det) through contract No.~638-2013-8993 and the Oskar Klein Centre for Cosmoparticle Physics. He also thanks The University of Texas at Austin for hospitality received during this work.

\appendix

\section{Appendices}

{\it A1.~Sensitivity to Primordial Non-Gaussianity and Model Dependence{\;---\;}}Non-Gaussian (NG) effects will change the probability of PBH formation and thus their dark matter contribution \cite{Bullock:1996at, Kalaja:2019uju}. There are many NG effects that modify the probability of collapse, from changes in the tail of the primordial density contrast distribution function to non-linear effects in the gravitational collapse at PBH formation. We emphasise that our model for the generation of curvature fluctuations on QCD scales is different from the multiple-field (curvaton) model presented in Refs.~\cite{Carr:2019hud, GarciaBellido:2019vlf}. Here we envision an inflation model with two slow-roll phases. For example, Critical Higgs Inflation \cite{Ezquiaga:2017fvi, Garcia-Bellido:2017mdw} may induce fluctuations of order $10^{-5}$ on the CMB scale and $0.1$ on the PBH scale. Both can be generated by dynamics consistent with present values of SM parameters and this gives similar spectral tilts ($\tilde{n}_{\srm} \sim 0.96$) on two very different scales. The assumption of near-scale-invariance can be relaxed to describe more complex formation mechanisms, but the thermal history will still imprint the PBH mass function in a similar way. These features are therefore universal and would apply for any PBH model.
 
Note that the $\Ocal( 1 )$ fluctuations needed for PBH collapse are generic in inflation \cite{Hawking:1982cz}. What is unusual is the small amplitude observed in the CMB, which requires some adjustment of parameters, although realised very naturally in Higgs inflation \cite{Ezquiaga:2017fvi}. Moreover, the NG change in the tail of the PDF of curvature fluctuations due to deviations from slow-roll at the quasi-inflection point in critical Higgs inflation will give rise to an exponential amplification of the probability of collapse to PBHs when those scales re-enter the Hubble scale during the radiation era \cite{Ezquiaga:2018gbw}. We have therefore assumed a nearly scale-invariant spectrum characterises this type of single-field model of inflation. A more detailed study will be required to test those scales against the observed mass distribution in the LIGO/Virgo events.

{\it A2.~Microlensing Limits{\;---\;}}
The most important uncertainty in the microlensing limits comes from PBH clustering. Rather than being uniformly distributed in the galactic halo, as assumed by the MACHO, EROS and OGLE analyses, PBHs must be clustered to some degree. This applies for all forms of dark matter but the clustering in the PBH case is strengthened for two reasons. First, the Poisson fluctuations due to the discrete nature of PBHs give a dominant contribution to the matter power spectrum below kiloparsec scales \cite{Carr:2018rid}. Our fourth cosmic conundrum relies on this effect. Second, the strongly enhanced and nearly scale-invariant primordial power spectrum ensures that \textit{any} small-scale inhomogeneity is rapidly driven into the non-linear regime. Even if the impact on the halo mass function is not so significant, overdense regions collapse at much earlier times and this leads to the formation of dark matter clusters (minihalos) which are initially more compact than in the standard cosmological scenario. 

The size of these PBH minihalos today is nevertheless limited by dynamical effects, just as for globular clusters. The typical relaxation time for a halo made of compact objects is 
\begin{AppEquation}
	t_{\rm rel}
		\approx
				\left(
					\frac{ N_{\rm cl} }
						{ \ln N_{\rm cl}}
				\right)
				\frac{ v_{\rm vir} }
					{ R_{\rm cl} }
				\, ,
\end{AppEquation}
where $N_{\rm cl}$, $R_{\rm cl}$ and $v_{\rm vir}$ are the number of PBHs in the minihalo, its radius and its virial velocity,
\begin{AppEquation}
	v_{\rm vir}
		\approx
				\sqrt{
					\frac{ G M_{\rm cl} }
					{ 2\.R_{\rm cl} }
				\,}
				\, ,
\end{AppEquation}
respectively. Assuming one requires $t_{\rm rel} > 10^{10}\,$yr for the clusters to survive and a mass $M_{\rm PBH} \approx 2\,\Msun$, corresponding to the proton peak in the PBH mass function, one gets a lower limit $R_{\rm cl} \approx 10\,$pc. The derivation of this limit can be found in Ref.~\cite{Clesse:2017bsw}. The dependence on the halo mass is only logarithmic and so this critical scale is universal.

A more refined analysis of the dynamics of UFDGs, including the effects of a central IMBH, leads to a similar lower limit on $R_{\rm cl}$ and is supported by UFDG observations (see the fifth cosmic conundrum). For globular clusters, the dynamical hardening of binaries in the core can reduce the critical radius to parsec scales and this might also apply for some PBH clusters. But it is unlikely that smaller PBH clusters exist today, because they would be dynamically unstable. 
 
Given that clustering is unavoidable, the main contribution to a uniform PBH distribution in our own galactic halo would come from PBHs which have been removed from their host cluster by close encounters and tidal interactions (\ie~the slingshot mechanism). Preliminary $N$-body simulations show that the fraction of objects ejected over the age of the Universe is about $10$\%. The case of PBHs is not much different and those with a mass comparable to stars should be expelled at a similar rate. For subsolar PBHs, the uniform fraction could be enhanced but they are anyway contributing no more than a few percents of the dark matter, much below the microlensing limit on non-clustered PBHs.

We now explain why PBHs in clusters do not induce detectable microlensing events, thereby evading the usual LMC and SMC microlensing limits, even if PBHs provide all the dark matter. If a PBH cluster, assumed to be spherical for simplicity, is aligned with a star in one of the Magellanic clouds, it will induce strong gravitational lensing of this star, with a deflection angle 
\begin{AppEquation}
	\alpha( \xi )
		=
				\frac{ 4\.G M( \xi ) }
				{ c^{2}\.\xi }
				\, ,
\end{AppEquation}
where 
\begin{AppEquation}
	M( \xi )
		=
				2 \pi
				\int_{0}^{\xi}\drm\xi'\;
				\Sigma( \xi' )\.\xi'
\end{AppEquation}
is the mass within a cylinder of radius $\xi$ with axis along the $z$ direction between the observer and the star, and 
\begin{AppEquation}
	\Sigma ( \xi )
		=
				\int\drm z\;
				\rho_{\rm PBH}( \xi, z )
\end{AppEquation}
is the projected surface density of the PBH cluster. For a cluster of mass $M_{\rm cl}$ and radius $R_{\rm cl}\approx 10\,$pc and $\xi \simeq R_{\rm cl}$, the deflection angle is
\begin{AppEquation}
	\alpha( R_{\rm cl}, M_{\rm cl} )
		\approx
				2 \times 10^{-13}
				\left(
					\frac{ M_{\rm cl} }
					{\,\Msun }
				\right)\!
				\left(
					\frac{ \rm pc }
					{ R_{\rm cl} }
				\right)
		\approx
				2 \times 10^{-10}
				\, .
\end{AppEquation}
Typically, this is not resolvable but the observed luminosity flux of the star is spread over an Einstein arc. At a cluster distance $D_{\rm cl} \sim \Ocal( 10\,{\rm kpc} )$, the deflection angle is subtended by a length $L \sim D_{\rm cl}\.\alpha \sim 10^{-9}\.( M_{\rm cl} /\,\Msun )$, which for a cluster mass $M_{\rm cl} \gtrsim 10^{3}\,\Msun$ is much larger than the Einstein radius $R_{\rm E}$ of an individual solar-mass PBH in the cluster:
\begin{AppEquation}
	R_{\rm E}
		=
				2\.
				\sqrt{
					G\.m^{}_{\rm PBH}\,
					x\.( 1 - x )\.
					\frac{ D_{\rm cl} }
					{ c^{2} }
				\,}
		\sim
				10^{-8}\,{\rm pc}
				\, ,
\end{AppEquation}
where $x \equiv D_{\rm cl} / D_{\rm LMC / SMC}$ is the distance ratio between the lens and the source. Therefore the star's luminosity is only marginally affected by magnification due to a nearby PBH. Indeed, the microlensing by a PBH in a cluster would induce less than $10$\% magnification of the star, whereas microlensing surveys only searched for $\Ocal( 1 )$ magnifications \cite{Fleury:2019xzr}. Only if the star is exactly aligned with the center of mass of the cluster, which is very unlikely, or if the cluster mass is below $10^{3}\,\Msun$, which should only represent only a small fraction of the PBH density, would the magnification of the star's luminosity be appreciable.

This new and rather simple argument provide robust motivation for the microlensing limits presented in line 4 of Table~1 and in Fig.~2. It also applies to the supernova \cite{Zumalacarregui:2017qqd, Garcia-Bellido:2017imq} and quasar microlensing limits, since the Einstein radius scales as $R_{\rm E} \propto D_{\rm cl}^{1/2}$ and the arc length as $L \propto D_{\rm cl}$, suppressing the deflection angle even more. Finally, invoking clustering to evade galactic and quasar microlensing limits does not contradict the PBH interpretation of OGLE microlensing events towards the Galactic center, because dense PBH clusters are probably tidally disrupted there, leading to a smoother PBH distribution. Nor does it affect our explanation of the second conundrum (quasar microlensing) since this only requires a homogeneous distribution with $f_{\rm PBH} \sim 0.01$--$0.1$, which is consistent with our clustering scenario.
\vs{1mm}

{\it A3.~Distribution of PBH Merger Detections{\;---\;}}The expected distribution of PBH merger detections as a function of the component masses $m_{1}$ and $m_{2}$ for LIGO/Virgo has been estimated for PBH binaries formed through tidal capture in dark matter halos. Their merging time $\tau$ is given by \cite{Clesse:2016vqa}
\begin{AppEquation}
	\frac{ \drm^{2} \tau }{ \drm m_{1} \drm m _{2} }
		\propto
				f_{\rm PBH}( m_{1} )\.
				f_{\rm PBH}( m_{2} )
				\times
				\frac{ ( m_{1}\.m_{2} )^{ 2 / 7 } }
				{ ( m_{1} + m_{2} )^{10 / 7} }
				\, .
				\label{eq:tau}
\end{AppEquation}
The normalisation is unimportant as long as one is interested in the \textit{distribution} of BH mergers. The total rate has uncertainties related to the halo mass function and concentration, the minihalo profiles, the PBH velocity distribution etc. Nevertheless, for realistic assumptions one can obtain a merging rate compatible with that inferred by LIGO/Virgo if PBHs constitute an appreciable fraction of the dark matter \cite{Bird:2016dcv, Clesse:2016vqa}. 

The \textit{distribution} of mergers as a function of the component masses is less impacted by these astrophysical uncertainties. It is therefore a better discriminant between BHs of primordial and stellar origin. Knowing the detector range as a function of the black hole or neutron star binary chirp mass \cite{Chen:2017wpg}, one can relate the rate distribution to the probability distribution for merger events. The latter is shown in Fig.~\ref{fig:fPBH} in the main text for the O2 observing run, assuming $n_{\srm} = 0.96$ and a PBH to horizon mass ratio $\gamma = 0.7$.

The expected distribution of merger \textit{detections} is the product of the comoving merging rate and the comoving volume probed by the detectors, $V_{\rm det} = ( 4\pi / 3 )\.R_{\rm det}^{3}$, where $R_{\rm det}$ is the luminosity distance of the furthest detectable source with a signal-to-noise ratio exceeding the threshold value of $8$. This `astrophysical range' depends on the detector sensitivity at the different frequencies of the gravitational wave (GW) signal. A simple estimate is given in Ref.~\cite{Chen:2017wpg} (and references therein), where different effects are analysed with more precise numerical computations. Here we use the simple estimate,
\begin{AppEquation}
	R_{\rm det}
		=
				\frac{ \sqrt{5\,} }
				{ 24 }\.
				\frac{( G \mathcal M c^{3})^{5/6}}
				{ \pi^{2/3} }
				\times
				\frac{ 1 }{ 2.26 }
				\left[
					\int_{f_{\rm min} }^{f_{\rm max}}\d f\;
					\frac{ f^{- \alpha} }
					{ S_{h}( f ) }
				\right]^{1/2}
				\, ,
				\label{eq:range}
\end{AppEquation}
where $\Mcal \equiv ( m_{1}\.m_{2} )^{3/5} / ( m_{1} + m_{2} )^{1/5}$ is the chirp mass, the factor $2.26$ comes from the ratio of the range to the horizon in a Euclidian Universe, and the integral is the contribution to the signal-to-noise ratio for the whole signal duration over the frequency bandwidth of the detector, assumed to have a noise power spectral density $S_{h}$.

\begin{figure}
	\vs{-1mm}
	\centering 
	\includegraphics[width = 0.42\textwidth]{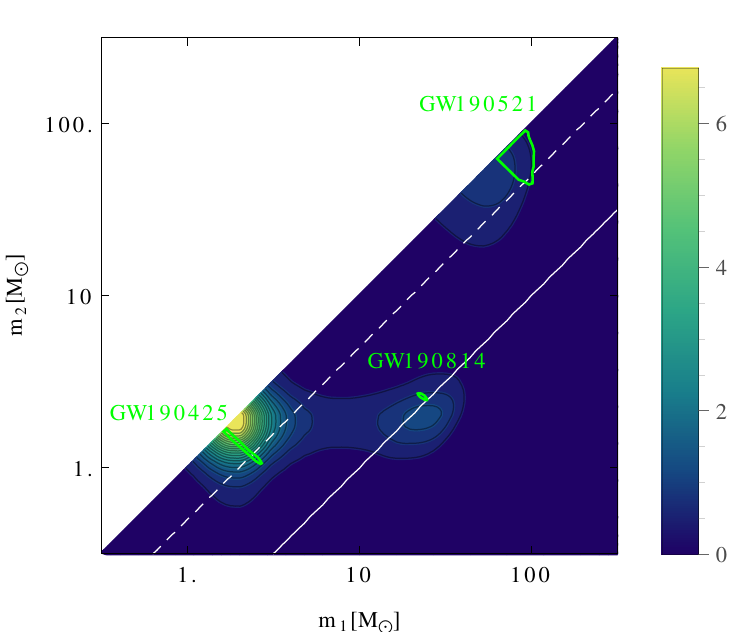}
	\caption{
			Expected probability distribution of PBH binary merger detections with masses 
			$m_{1}$ and $m_{2}$ (in units of solar mass)
			by LIGO/Virgo for the primordial binary formation channel, 
			using the same method and PBH mass function as
			Fig.~\ref{fig:fPBH} in the main text.
			The merger rate distribution is provided in \cite{Gow:2019pok} and 
			we assume a suppression factor $S = 0.01$ to get a total merging rate of 
			$\tau \approx 50 {\rm yr}^{-1} {\rm Gpc}^{-3}$ 
			between $5$ and $50\,\Msun$.
			Predominant mergers in the \textit{proton peak} are expected for this channel, 
			which is excluded by the mass distribution of O1 and O2 detections.
			The solid and dashed white lines correspond to mass ratios 
			$q = m_{2} / m_{1}$ of $0.1$ and $0.5$, respectively.
			\vs{-6mm}}
	\label{fig:Rate-prim}
\end{figure}

For a simple post-Newtonian waveform of the inspiral phase, one has $\alpha = 7 / 3$ until the GW frequency associated with the innermost stable circular orbit (ISCO),
\begin{AppEquation}
	f_{\rm ISCO}( m_{1}, m_{2} )
		=
				\frac{ 4400\,{\rm Hz} }
				{ ( m_{1} + m_{2} ) }
				\times
				\frac{ 4 \, m_{1}\.m_{2} }
				{ ( m_{1} + m_{2} )^{2} }
				\, ,
\end{AppEquation}
where the black hole masses are in solar units. However, in our scenario numerous stellar-mass black holes from the peak of the mass function can be captured by heavier black holes from the pion bump thereby producing binaries with mass ratios of order 0.1, so it is crucial to take this effect into account. Typically, it reduces the astrophysical range and hence detection rate for low-mass ratios but there is still a significant probability of detection in the interesting region 5 of Fig.~\ref{fig:Rate-prim} where GW190814 has been observed. We also include the merging phase in our analysis, with $\alpha = 2/3$ between $f_{\rm ISCO}$ and the merger frequency
\begin{equation}
	f_{\rm merge}( m_{1}, m_{2} )
		=
				2\.f_{\rm ISCO}
				\times
				\frac{ 4 \, m_{1}\.m_{2} }
				{ ( m_{1} + m_{2} )^{2} }
				\, .
\end{equation}
The last factor can be found in Ref.~\cite{Lousto:2020tnb} and is around one for mass ratios close to unity. We take the lower and upper bounds of the integral to be $f_{\rm min}= 50\,{\rm Hz}$ and $f_{\rm max} = \min( f_{\rm merge}, 2000\,{\rm Hz} )$. Finally, we account for redshift effects on the comoving rates and waveform by replacing $\Mcal$ by $\Mcal( 1 + z )$ in Eq.~\eqref{eq:range}. We have used the detector noise power spectral density $S_{h}$ adopted by the LIGO-Livingstone detector~\cite{TheLIGOScientific:2017qsa} in the last month of its second observing run.

This approach still has some inaccuracies. The detector range is wrong by a factor of a few, so this has been corrected by a linear rescaling in order for the range to be in reasonable agreement with Fig.~4 of Ref.~\cite{Chen:2017wpg} and a binary neutron star range of $90$\,Mpc~\cite{Abbott:2019ebz}. In analysing the third observing run, a more accurate approach may be adopted by carrying out campaigns over the whole run with different sky locations and by using more accurate GW waveforms during the merging and ring-down phases. However, such an analysis is beyond the scope of this paper, in which we estimate the expected event distribution for the second LIGO/Virgo observing, using public data of the detector noise spectral density. The comparison with some events observed in the third observing run is thus only qualitative.

So far the expected event distribution is compatible with observation and its specific features are discussed in the main text. We do not consider the expected event distribution of detections for primordial binaries because it is still unclear how the expected PBH mass function would suppress the merging rate due to early-forming PBH halos seeded by Poisson fluctuations. A more refined analysis of this binary formation channel is needed and this will require $N$-body simulations.

\vspace{1mm}

{\it A4.~Late versus early binary formation{\;---\;}}It is still unclear which binary-formation channel dominates. However, the primordial channel is subject to more uncertainties and there are two reasons for assuming this is disfavoured. First, we are considering a scenario with $f_{\rm PBH} = 1$, in which case $N$-body simulations show that rapidly forming clusters lead to a suppression of the merging rate of primordial binaries for both a monochromatic and lognormal mass distribution. The importance of this suppression is still unclear but Ref.~\cite{Vaskonen:2019jpv} claims it could be up to two orders of magnitude. In our scenario, PBHs in the range $10$--$50\,\Msun$ relevant for LIGO/Virgo account for only a few percent of the dark matter. The two effects combine and the merging rate becomes of order $1$--$10\,$yr$^{-1}$\,Gpc$^{-3}$, consistent with the LIGO/Virgo lower bound. Consequently, the merging rate from captures in halos can dominate. Since the mass dependence of the suppression for primordial binaries is unknown for our mass function, we have only considered the capture channel because the merger rate distribution is subject to fewer uncertainties in this case.

Second, the rate limits from the O1--\.O2 runs of LIGO/Virgo already disfavour primordial binaries as the dominant channel for our mass distribution if we assume (for simplicity) a suppression rate independent of the mass or even no suppression at all. The merging rates should go roughly as $f( m_{\rm PBH} )^{2} / m_{\rm PBH}$ for equal-mass binaries. For unequal-mass binaries, the rates calculated by Kocsis {\it et al.} \cite{Kocsis:2017yty} and assumed by Gow {\it et al.} \cite{Gow:2019pok} can be used but one comes to the same conclusion: with our bumpy mass function the limits set by LIGO/Virgo on the subsolar scale cannot be evaded for primordial binaries if they also explain some of the larger black hole mergers. We have also reproduced Fig.~\ref{fig:fPBH} for the case of primordial binaries in Fig.~\ref{fig:Rate-prim}. The expected distribution of merger events in the ($m_{1},\.m_{2}$) plane has a dominant peak at the solar-mass scale and only a subdominant number of events the high-mass region, which is the opposite of what LIGO/Virgo has observed. Finally, we point out the recent claim of Ref.~\cite{Boehm:2020jwd} that the expected merging rate of primordial binaries is highly suppressed due to the time dependence of the Misner-Sharp mass in the Thakurta metric for the radial density flows.

\bibliography{refs}

\begin{thebibliography}{90}%
\makeatletter
\providecommand \@ifxundefined [1]{%
 \@ifx{#1\undefined}
}%
\providecommand \@ifnum [1]{%
 \ifnum #1\expandafter \@firstoftwo
 \else \expandafter \@secondoftwo
 \fi
}%
\providecommand \@ifx [1]{%
 \ifx #1\expandafter \@firstoftwo
 \else \expandafter \@secondoftwo
 \fi
}%
\providecommand \natexlab [1]{#1}%
\providecommand \enquote  [1]{``#1''}%
\providecommand \bibnamefont  [1]{#1}%
\providecommand \bibfnamefont [1]{#1}%
\providecommand \citenamefont [1]{#1}%
\providecommand \href@noop [0]{\@secondoftwo}%
\providecommand \href [0]{\begingroup \@sanitize@url \@href}%
\providecommand \@href[1]{\@@startlink{#1}\@@href}%
\providecommand \@@href[1]{\endgroup#1\@@endlink}%
\providecommand \@sanitize@url [0]{\catcode `\\12\catcode `\$12\catcode
  `\&12\catcode `\#12\catcode `\^12\catcode `\_12\catcode `\%12\relax}%
\providecommand \@@startlink[1]{}%
\providecommand \@@endlink[0]{}%
\providecommand \url  [0]{\begingroup\@sanitize@url \@url }%
\providecommand \@url [1]{\endgroup\@href {#1}{\urlprefix }}%
\providecommand \urlprefix  [0]{URL }%
\providecommand \Eprint [0]{\href }%
\providecommand \doibase [0]{http://dx.doi.org/}%
\providecommand \selectlanguage [0]{\@gobble}%
\providecommand \bibinfo  [0]{\@secondoftwo}%
\providecommand \bibfield  [0]{\@secondoftwo}%
\providecommand \translation [1]{[#1]}%
\providecommand \BibitemOpen [0]{}%
\providecommand \bibitemStop [0]{}%
\providecommand \bibitemNoStop [0]{.\EOS\space}%
\providecommand \EOS [0]{\spacefactor3000\relax}%
\providecommand \BibitemShut  [1]{\csname bibitem#1\endcsname}%
\let\auto@bib@innerbib\@empty
\bibitem [{\citenamefont {Abbott}\ \emph {et~al.}(2016)\citenamefont {Abbott}
  \emph {et~al.}}]{Abbott:2016nmj}%
  \BibitemOpen
  \bibfield  {author} {\bibinfo {author} {\bibfnamefont {B.~P.}\ \bibnamefont
  {Abbott}} \emph {et~al.} (\bibinfo {collaboration} {Virgo, LIGO Scientific
  Collaborations}),\ }\href {\doibase 10.1103/PhysRevLett.116.241103}
  {\bibfield  {journal} {\bibinfo  {journal} {Phys.~Rev.~Lett.}\ }\textbf
  {\bibinfo {volume} {116}},\ \bibinfo {pages} {241103} (\bibinfo {year}
  {2016})}\BibitemShut {NoStop}%
\bibitem [{\citenamefont {Wysocki}\ \emph {et~al.}(2018)\citenamefont
  {Wysocki}, \citenamefont {Gerosa}, \citenamefont {O'Shaughnessy},
  \citenamefont {Belczynski}, \citenamefont {Gladysz}, \citenamefont {Berti},
  \citenamefont {Kesden},\ and\ \citenamefont {Holz}}]{Wysocki:2017isg}%
  \BibitemOpen
  \bibfield  {author} {\bibinfo {author} {\bibfnamefont {D.}~\bibnamefont
  {Wysocki}}, \bibinfo {author} {\bibfnamefont {D.}~\bibnamefont {Gerosa}},
  \bibinfo {author} {\bibfnamefont {R.}~\bibnamefont {O'Shaughnessy}}, \bibinfo
  {author} {\bibfnamefont {K.}~\bibnamefont {Belczynski}}, \bibinfo {author}
  {\bibfnamefont {W.}~\bibnamefont {Gladysz}}, \bibinfo {author} {\bibfnamefont
  {E.}~\bibnamefont {Berti}}, \bibinfo {author} {\bibfnamefont
  {M.}~\bibnamefont {Kesden}}, \ and\ \bibinfo {author} {\bibfnamefont {D.~E.}\
  \bibnamefont {Holz}},\ }\href {\doibase 10.1103/PhysRevD.97.043014}
  {\bibfield  {journal} {\bibinfo  {journal} {Phys. Rev.}\ }\textbf {\bibinfo
  {volume} {D97}},\ \bibinfo {pages} {043014} (\bibinfo {year} {2018})},\
  \Eprint {http://arxiv.org/abs/1709.01943} {arXiv:1709.01943 [astro-ph.HE]}
  \BibitemShut {NoStop}%
\bibitem [{\citenamefont {Garc{\'i}a-Bellido}\ \emph
  {et~al.}(1996)\citenamefont {Garc{\'i}a-Bellido}, \citenamefont {Linde},\
  and\ \citenamefont {Wands}}]{GarciaBellido:1996qt}%
  \BibitemOpen
  \bibfield  {author} {\bibinfo {author} {\bibfnamefont {J.}~\bibnamefont
  {Garc{\'i}a-Bellido}}, \bibinfo {author} {\bibfnamefont {A.~D.}\ \bibnamefont
  {Linde}}, \ and\ \bibinfo {author} {\bibfnamefont {D.}~\bibnamefont
  {Wands}},\ }\href {\doibase 10.1103/PhysRevD.54.6040} {\bibfield  {journal}
  {\bibinfo  {journal} {Phys. Rev.}\ }\textbf {\bibinfo {volume} {D54}},\
  \bibinfo {pages} {6040} (\bibinfo {year} {1996})},\ \Eprint
  {http://arxiv.org/abs/astro-ph/9605094} {arXiv:astro-ph/9605094 [astro-ph]}
  \BibitemShut {NoStop}%
\bibitem [{\citenamefont {Carr}(1975)}]{Carr:1975qj}%
  \BibitemOpen
  \bibfield  {author} {\bibinfo {author} {\bibfnamefont {B.~J.}\ \bibnamefont
  {Carr}},\ }\href {\doibase 10.1086/153853} {\bibfield  {journal} {\bibinfo
  {journal} {Astrophys.~J.}\ }\textbf {\bibinfo {volume} {201}},\ \bibinfo
  {pages} {1} (\bibinfo {year} {1975})}\BibitemShut {NoStop}%
\bibitem [{\citenamefont {Schmid}\ \emph {et~al.}(1999)\citenamefont {Schmid},
  \citenamefont {Schwarz},\ and\ \citenamefont {Widerin}}]{Schmid:1998mx}%
  \BibitemOpen
  \bibfield  {author} {\bibinfo {author} {\bibfnamefont {C.}~\bibnamefont
  {Schmid}}, \bibinfo {author} {\bibfnamefont {D.~J.}\ \bibnamefont {Schwarz}},
  \ and\ \bibinfo {author} {\bibfnamefont {P.}~\bibnamefont {Widerin}},\ }\href
  {\doibase 10.1103/PhysRevD.59.043517} {\bibfield  {journal} {\bibinfo
  {journal} {Phys.~Rev.}\ }\textbf {\bibinfo {volume} {D59}},\ \bibinfo {pages}
  {043517} (\bibinfo {year} {1999})},\ \Eprint
  {http://arxiv.org/abs/astro-ph/9807257} {arXiv:astro-ph/9807257 [astro-ph]}
  \BibitemShut {NoStop}%
\bibitem [{\citenamefont {Laine}\ and\ \citenamefont
  {Schroder}(2006)}]{Laine:2006cp}%
  \BibitemOpen
  \bibfield  {author} {\bibinfo {author} {\bibfnamefont {M.}~\bibnamefont
  {Laine}}\ and\ \bibinfo {author} {\bibfnamefont {Y.}~\bibnamefont
  {Schroder}},\ }\href {\doibase 10.1103/PhysRevD.73.085009} {\bibfield
  {journal} {\bibinfo  {journal} {Phys. Rev.}\ }\textbf {\bibinfo {volume}
  {D73}},\ \bibinfo {pages} {085009} (\bibinfo {year} {2006})},\ \Eprint
  {http://arxiv.org/abs/hep-ph/0603048} {arXiv:hep-ph/0603048 [hep-ph]}
  \BibitemShut {NoStop}%
\bibitem [{\citenamefont {Bhattacharya}\ \emph {et~al.}(2014)\citenamefont
  {Bhattacharya} \emph {et~al.}}]{Bhattacharya:2014ara}%
  \BibitemOpen
  \bibfield  {author} {\bibinfo {author} {\bibfnamefont {T.}~\bibnamefont
  {Bhattacharya}} \emph {et~al.},\ }\href {\doibase
  10.1103/PhysRevLett.113.082001} {\bibfield  {journal} {\bibinfo  {journal}
  {Phys. Rev. Lett.}\ }\textbf {\bibinfo {volume} {113}},\ \bibinfo {pages}
  {082001} (\bibinfo {year} {2014})},\ \Eprint {http://arxiv.org/abs/1402.5175}
  {arXiv:1402.5175 [hep-lat]} \BibitemShut {NoStop}%
\bibitem [{\citenamefont {Laine}\ and\ \citenamefont
  {Meyer}(2015)}]{Laine:2015kra}%
  \BibitemOpen
  \bibfield  {author} {\bibinfo {author} {\bibfnamefont {M.}~\bibnamefont
  {Laine}}\ and\ \bibinfo {author} {\bibfnamefont {M.}~\bibnamefont {Meyer}},\
  }\href {\doibase 10.1088/1475-7516/2015/07/035} {\bibfield  {journal}
  {\bibinfo  {journal} {JCAP}\ }\textbf {\bibinfo {volume} {07}},\ \bibinfo
  {pages} {035} (\bibinfo {year} {2015})},\ \Eprint
  {http://arxiv.org/abs/1503.04935} {arXiv:1503.04935 [hep-ph]} \BibitemShut
  {NoStop}%
\bibitem [{\citenamefont {Borsanyi}\ \emph {et~al.}(2016)\citenamefont
  {Borsanyi} \emph {et~al.}}]{Borsanyi:2016ksw}%
  \BibitemOpen
  \bibfield  {author} {\bibinfo {author} {\bibfnamefont {S.}~\bibnamefont
  {Borsanyi}} \emph {et~al.},\ }\href {\doibase 10.1038/nature20115} {\bibfield
   {journal} {\bibinfo  {journal} {Nature}\ }\textbf {\bibinfo {volume}
  {539}},\ \bibinfo {pages} {69} (\bibinfo {year} {2016})},\ \Eprint
  {http://arxiv.org/abs/1606.07494} {arXiv:1606.07494 [hep-lat]} \BibitemShut
  {NoStop}%
\bibitem [{\citenamefont {Byrnes}\ \emph {et~al.}(2018)\citenamefont {Byrnes},
  \citenamefont {Hindmarsh}, \citenamefont {Young},\ and\ \citenamefont
  {Hawkins}}]{Byrnes:2018clq}%
  \BibitemOpen
  \bibfield  {author} {\bibinfo {author} {\bibfnamefont {C.~T.}\ \bibnamefont
  {Byrnes}}, \bibinfo {author} {\bibfnamefont {M.}~\bibnamefont {Hindmarsh}},
  \bibinfo {author} {\bibfnamefont {S.}~\bibnamefont {Young}}, \ and\ \bibinfo
  {author} {\bibfnamefont {M.~R.~S.}\ \bibnamefont {Hawkins}},\ }\href
  {\doibase 10.1088/1475-7516/2018/08/041} {\bibfield  {journal} {\bibinfo
  {journal} {JCAP}\ }\textbf {\bibinfo {volume} {08}},\ \bibinfo {pages} {041}
  (\bibinfo {year} {2018})},\ \Eprint {http://arxiv.org/abs/1801.06138}
  {arXiv:1801.06138 [astro-ph.CO]} \BibitemShut {NoStop}%
\bibitem [{\citenamefont {Saikawa}\ and\ \citenamefont
  {Shirai}(2018)}]{Saikawa:2018rcs}%
  \BibitemOpen
  \bibfield  {author} {\bibinfo {author} {\bibfnamefont {K.}~\bibnamefont
  {Saikawa}}\ and\ \bibinfo {author} {\bibfnamefont {S.}~\bibnamefont
  {Shirai}},\ }\href {\doibase 10.1088/1475-7516/2018/05/035} {\bibfield
  {journal} {\bibinfo  {journal} {JCAP}\ }\textbf {\bibinfo {volume} {05}},\
  \bibinfo {pages} {035} (\bibinfo {year} {2018})},\ \Eprint
  {http://arxiv.org/abs/1803.01038} {arXiv:1803.01038 [hep-ph]} \BibitemShut
  {NoStop}%
\bibitem [{\citenamefont {Carr}\ \emph {et~al.}(2019)\citenamefont {Carr},
  \citenamefont {Clesse},\ and\ \citenamefont
  {Garc{\'i}a-Bellido}}]{Carr:2019hud}%
  \BibitemOpen
  \bibfield  {author} {\bibinfo {author} {\bibfnamefont {B.}~\bibnamefont
  {Carr}}, \bibinfo {author} {\bibfnamefont {S.}~\bibnamefont {Clesse}}, \ and\
  \bibinfo {author} {\bibfnamefont {J.}~\bibnamefont {Garc{\'i}a-Bellido}},\
  }\href@noop {} {\  (\bibinfo {year} {2019})},\ \Eprint
  {http://arxiv.org/abs/1904.02129} {arXiv:1904.02129 [astro-ph.CO]}
  \BibitemShut {NoStop}%
\bibitem [{\citenamefont {Garc{\'i}a-Bellido}\ \emph
  {et~al.}(2019)\citenamefont {Garc{\'i}a-Bellido}, \citenamefont {Carr},\ and\
  \citenamefont {Clesse}}]{GarciaBellido:2019vlf}%
  \BibitemOpen
  \bibfield  {author} {\bibinfo {author} {\bibfnamefont {J.}~\bibnamefont
  {Garc{\'i}a-Bellido}}, \bibinfo {author} {\bibfnamefont {B.}~\bibnamefont
  {Carr}}, \ and\ \bibinfo {author} {\bibfnamefont {S.}~\bibnamefont
  {Clesse}},\ }\href@noop {} {\  (\bibinfo {year} {2019})},\ \Eprint
  {http://arxiv.org/abs/1904.11482} {arXiv:1904.11482 [astro-ph.CO]}
  \BibitemShut {NoStop}%
\bibitem [{\citenamefont {Niikura}\ \emph {et~al.}(2019)\citenamefont
  {Niikura}, \citenamefont {Takada}, \citenamefont {Yokoyama}, \citenamefont
  {Sumi},\ and\ \citenamefont {Masaki}}]{Niikura:2019kqi}%
  \BibitemOpen
  \bibfield  {author} {\bibinfo {author} {\bibfnamefont {H.}~\bibnamefont
  {Niikura}}, \bibinfo {author} {\bibfnamefont {M.}~\bibnamefont {Takada}},
  \bibinfo {author} {\bibfnamefont {S.}~\bibnamefont {Yokoyama}}, \bibinfo
  {author} {\bibfnamefont {T.}~\bibnamefont {Sumi}}, \ and\ \bibinfo {author}
  {\bibfnamefont {S.}~\bibnamefont {Masaki}},\ }\href {\doibase
  10.1103/PhysRevD.99.083503} {\bibfield  {journal} {\bibinfo  {journal} {Phys.
  Rev.}\ }\textbf {\bibinfo {volume} {D99}},\ \bibinfo {pages} {083503}
  (\bibinfo {year} {2019})},\ \Eprint {http://arxiv.org/abs/1901.07120}
  {arXiv:1901.07120 [astro-ph.CO]} \BibitemShut {NoStop}%
\bibitem [{\citenamefont {Mediavilla}\ \emph {et~al.}(2017)\citenamefont
  {Mediavilla}, \citenamefont {Jim\'enez-Vicente}, \citenamefont {Mu\~noz},
  \citenamefont {Vives-Arias},\ and\ \citenamefont
  {Calder\'on-Infante}}]{Mediavilla:2017bok}%
  \BibitemOpen
  \bibfield  {author} {\bibinfo {author} {\bibfnamefont {E.}~\bibnamefont
  {Mediavilla}}, \bibinfo {author} {\bibfnamefont {J.}~\bibnamefont
  {Jim\'enez-Vicente}}, \bibinfo {author} {\bibfnamefont {J.~A.}\ \bibnamefont
  {Mu\~noz}}, \bibinfo {author} {\bibfnamefont {H.}~\bibnamefont
  {Vives-Arias}}, \ and\ \bibinfo {author} {\bibfnamefont {J.}~\bibnamefont
  {Calder\'on-Infante}},\ }\href {\doibase 10.3847/2041-8213/aa5dab} {\bibfield
   {journal} {\bibinfo  {journal} {Astrophys. J.}\ }\textbf {\bibinfo {volume}
  {836}},\ \bibinfo {pages} {L18} (\bibinfo {year} {2017})},\ \Eprint
  {http://arxiv.org/abs/1702.00947} {arXiv:1702.00947 [astro-ph.GA]}
  \BibitemShut {NoStop}%
\bibitem [{\citenamefont {Wyrzykowski}\ and\ \citenamefont
  {Mandel}(2019)}]{Wyrzykowski:2019jyg}%
  \BibitemOpen
  \bibfield  {author} {\bibinfo {author} {\bibfnamefont {L.}~\bibnamefont
  {Wyrzykowski}}\ and\ \bibinfo {author} {\bibfnamefont {I.}~\bibnamefont
  {Mandel}},\ }\href@noop {} {\  (\bibinfo {year} {2019})},\ \Eprint
  {http://arxiv.org/abs/1904.07789} {arXiv:1904.07789 [astro-ph.SR]}
  \BibitemShut {NoStop}%
\bibitem [{\citenamefont {Brown}\ \emph {et~al.}(1999)\citenamefont {Brown},
  \citenamefont {Lee},\ and\ \citenamefont {Bethe}}]{Brown:1999ax}%
  \BibitemOpen
  \bibfield  {author} {\bibinfo {author} {\bibfnamefont {G.~E.}\ \bibnamefont
  {Brown}}, \bibinfo {author} {\bibfnamefont {C.~H.}\ \bibnamefont {Lee}}, \
  and\ \bibinfo {author} {\bibfnamefont {H.~A.}\ \bibnamefont {Bethe}},\
  }\href@noop {} {\  (\bibinfo {year} {1999})},\ \Eprint
  {http://arxiv.org/abs/astro-ph/9909270} {arXiv:astro-ph/9909270 [astro-ph]}
  \BibitemShut {NoStop}%
\bibitem [{\citenamefont {{Kashlinsky}}\ \emph {et~al.}(2005)\citenamefont
  {{Kashlinsky}}, \citenamefont {{Arendt}}, \citenamefont {{Mather}},\ and\
  \citenamefont {{Moseley}}}]{2005Natur.438...45K}%
  \BibitemOpen
  \bibfield  {author} {\bibinfo {author} {\bibfnamefont {A.}~\bibnamefont
  {{Kashlinsky}}}, \bibinfo {author} {\bibfnamefont {R.~G.}\ \bibnamefont
  {{Arendt}}}, \bibinfo {author} {\bibfnamefont {J.}~\bibnamefont {{Mather}}},
  \ and\ \bibinfo {author} {\bibfnamefont {S.~H.}\ \bibnamefont {{Moseley}}},\
  }\href {\doibase 10.1038/nature04143} {\bibfield  {journal} {\bibinfo
  {journal} {\nat}\ }\textbf {\bibinfo {volume} {438}},\ \bibinfo {pages} {45}
  (\bibinfo {year} {2005})},\ \Eprint {http://arxiv.org/abs/astro-ph/0511105}
  {astro-ph/0511105} \BibitemShut {NoStop}%
\bibitem [{\citenamefont {Clesse}\ and\ \citenamefont
  {Garc{\'i}a-Bellido}(2018)}]{Clesse:2017bsw}%
  \BibitemOpen
  \bibfield  {author} {\bibinfo {author} {\bibfnamefont {S.}~\bibnamefont
  {Clesse}}\ and\ \bibinfo {author} {\bibfnamefont {J.}~\bibnamefont
  {Garc{\'i}a-Bellido}},\ }\href {\doibase 10.1016/j.dark.2018.08.004}
  {\bibfield  {journal} {\bibinfo  {journal} {Phys. Dark Univ.}\ }\textbf
  {\bibinfo {volume} {22}},\ \bibinfo {pages} {137} (\bibinfo {year} {2018})},\
  \Eprint {http://arxiv.org/abs/1711.10458} {arXiv:1711.10458 [astro-ph.CO]}
  \BibitemShut {NoStop}%
\bibitem [{\citenamefont {Abbott}\ \emph {et~al.}(2018)\citenamefont {Abbott}
  \emph {et~al.}}]{LIGOScientific:2018mvr}%
  \BibitemOpen
  \bibfield  {author} {\bibinfo {author} {\bibfnamefont {B.~P.}\ \bibnamefont
  {Abbott}} \emph {et~al.} (\bibinfo {collaboration} {LIGO Scientific,
  Virgo}),\ }\href@noop {} {\  (\bibinfo {year} {2018})},\ \Eprint
  {http://arxiv.org/abs/1811.12907} {arXiv:1811.12907 [astro-ph.HE]}
  \BibitemShut {NoStop}%
\bibitem [{\citenamefont {Vovchenko}\ \emph {et~al.}(2020)\citenamefont
  {Vovchenko}, \citenamefont {Brandt}, \citenamefont {Cuteri}, \citenamefont
  {Endr\H{o}di}, \citenamefont {Hajkarim},\ and\ \citenamefont
  {Schaffner-Bielich}}]{Vovchenko:2020crk}%
  \BibitemOpen
  \bibfield  {author} {\bibinfo {author} {\bibfnamefont {V.}~\bibnamefont
  {Vovchenko}}, \bibinfo {author} {\bibfnamefont {B.~B.}\ \bibnamefont
  {Brandt}}, \bibinfo {author} {\bibfnamefont {F.}~\bibnamefont {Cuteri}},
  \bibinfo {author} {\bibfnamefont {G.}~\bibnamefont {Endr\H{o}di}}, \bibinfo
  {author} {\bibfnamefont {F.}~\bibnamefont {Hajkarim}}, \ and\ \bibinfo
  {author} {\bibfnamefont {J.}~\bibnamefont {Schaffner-Bielich}},\ }\href@noop
  {} {\  (\bibinfo {year} {2020})},\ \Eprint {http://arxiv.org/abs/2009.02309}
  {arXiv:2009.02309 [hep-ph]} \BibitemShut {NoStop}%
\bibitem [{\citenamefont {Oldengott}\ and\ \citenamefont
  {Schwarz}(2017)}]{Oldengott:2017tzj}%
  \BibitemOpen
  \bibfield  {author} {\bibinfo {author} {\bibfnamefont {I.~M.}\ \bibnamefont
  {Oldengott}}\ and\ \bibinfo {author} {\bibfnamefont {D.~J.}\ \bibnamefont
  {Schwarz}},\ }\href {\doibase 10.1209/0295-5075/119/29001} {\bibfield
  {journal} {\bibinfo  {journal} {EPL}\ }\textbf {\bibinfo {volume} {119}},\
  \bibinfo {pages} {29001} (\bibinfo {year} {2017})},\ \Eprint
  {http://arxiv.org/abs/1706.01705} {arXiv:1706.01705 [astro-ph.CO]}
  \BibitemShut {NoStop}%
\bibitem [{\citenamefont {Middeldorf-Wygas}\ \emph {et~al.}(2020)\citenamefont
  {Middeldorf-Wygas}, \citenamefont {Oldengott}, \citenamefont {B{\"o}deker},\
  and\ \citenamefont {Schwarz}}]{Middeldorf-Wygas:2020glx}%
  \BibitemOpen
  \bibfield  {author} {\bibinfo {author} {\bibfnamefont {M.~M.}\ \bibnamefont
  {Middeldorf-Wygas}}, \bibinfo {author} {\bibfnamefont {I.~M.}\ \bibnamefont
  {Oldengott}}, \bibinfo {author} {\bibfnamefont {D.}~\bibnamefont
  {B{\"o}deker}}, \ and\ \bibinfo {author} {\bibfnamefont {D.~J.}\ \bibnamefont
  {Schwarz}},\ }\href@noop {} {\  (\bibinfo {year} {2020})},\ \Eprint
  {http://arxiv.org/abs/2009.00036} {arXiv:2009.00036 [hep-ph]} \BibitemShut
  {NoStop}%
\bibitem [{\citenamefont {Bodeker}\ \emph {et~al.}(2020)\citenamefont
  {Bodeker}, \citenamefont {Kuhnel}, \citenamefont {Oldengott},\ and\
  \citenamefont {Schwarz}}]{Bodeker:2020stj}%
  \BibitemOpen
  \bibfield  {author} {\bibinfo {author} {\bibfnamefont {D.}~\bibnamefont
  {Bodeker}}, \bibinfo {author} {\bibfnamefont {F.}~\bibnamefont {Kuhnel}},
  \bibinfo {author} {\bibfnamefont {I.~M.}\ \bibnamefont {Oldengott}}, \ and\
  \bibinfo {author} {\bibfnamefont {D.~J.}\ \bibnamefont {Schwarz}},\
  }\href@noop {} {\  (\bibinfo {year} {2020})},\ \Eprint
  {http://arxiv.org/abs/2011.07283} {arXiv:2011.07283 [astro-ph.CO]}
  \BibitemShut {NoStop}%
\bibitem [{\citenamefont {Carr}\ and\ \citenamefont
  {K{\"u}hnel}(2020)}]{Carr:2020xqk}%
  \BibitemOpen
  \bibfield  {author} {\bibinfo {author} {\bibfnamefont {B.}~\bibnamefont
  {Carr}}\ and\ \bibinfo {author} {\bibfnamefont {F.}~\bibnamefont
  {K{\"u}hnel}},\ }\href {\doibase 10.1146/annurev-nucl-050520-125911} {\
  (\bibinfo {year} {2020}),\ 10.1146/annurev-nucl-050520-125911},\ \Eprint
  {http://arxiv.org/abs/2006.02838} {arXiv:2006.02838 [astro-ph.CO]}
  \BibitemShut {NoStop}%
\bibitem [{\citenamefont {Musco}\ and\ \citenamefont
  {Miller}(2013)}]{Musco:2012au}%
  \BibitemOpen
  \bibfield  {author} {\bibinfo {author} {\bibfnamefont {I.}~\bibnamefont
  {Musco}}\ and\ \bibinfo {author} {\bibfnamefont {J.~C.}\ \bibnamefont
  {Miller}},\ }\href {\doibase 10.1088/0264-9381/30/14/145009} {\bibfield
  {journal} {\bibinfo  {journal} {Classical Quantum Gravity}\ }\textbf
  {\bibinfo {volume} {30}},\ \bibinfo {pages} {145009} (\bibinfo {year}
  {2013})}\BibitemShut {NoStop}%
\bibitem [{\citenamefont {Musco}\ \emph {et~al.}(2020)\citenamefont {Musco},
  \citenamefont {De~Luca}, \citenamefont {Franciolini},\ and\ \citenamefont
  {Riotto}}]{Musco:2020jjb}%
  \BibitemOpen
  \bibfield  {author} {\bibinfo {author} {\bibfnamefont {I.}~\bibnamefont
  {Musco}}, \bibinfo {author} {\bibfnamefont {V.}~\bibnamefont {De~Luca}},
  \bibinfo {author} {\bibfnamefont {G.}~\bibnamefont {Franciolini}}, \ and\
  \bibinfo {author} {\bibfnamefont {A.}~\bibnamefont {Riotto}},\ }\href@noop {}
  {\  (\bibinfo {year} {2020})},\ \Eprint {http://arxiv.org/abs/2011.03014}
  {arXiv:2011.03014 [astro-ph.CO]} \BibitemShut {NoStop}%
\bibitem [{\citenamefont {Young}\ \emph {et~al.}(2014)\citenamefont {Young},
  \citenamefont {Byrnes},\ and\ \citenamefont {Sasaki}}]{Young:2014ana}%
  \BibitemOpen
  \bibfield  {author} {\bibinfo {author} {\bibfnamefont {S.}~\bibnamefont
  {Young}}, \bibinfo {author} {\bibfnamefont {C.~T.}\ \bibnamefont {Byrnes}}, \
  and\ \bibinfo {author} {\bibfnamefont {M.}~\bibnamefont {Sasaki}},\ }\href
  {\doibase 10.1088/1475-7516/2014/07/045} {\bibfield  {journal} {\bibinfo
  {journal} {JCAP}\ }\textbf {\bibinfo {volume} {1407}},\ \bibinfo {pages}
  {045} (\bibinfo {year} {2014})},\ \Eprint {http://arxiv.org/abs/1405.7023}
  {arXiv:1405.7023 [gr-qc]} \BibitemShut {NoStop}%
\bibitem [{\citenamefont {Yoo}\ \emph {et~al.}(2018)\citenamefont {Yoo},
  \citenamefont {Harada}, \citenamefont {Garriga},\ and\ \citenamefont
  {Kohri}}]{Yoo:2018kvb}%
  \BibitemOpen
  \bibfield  {author} {\bibinfo {author} {\bibfnamefont {C.-M.}\ \bibnamefont
  {Yoo}}, \bibinfo {author} {\bibfnamefont {T.}~\bibnamefont {Harada}},
  \bibinfo {author} {\bibfnamefont {J.}~\bibnamefont {Garriga}}, \ and\
  \bibinfo {author} {\bibfnamefont {K.}~\bibnamefont {Kohri}},\ }\href
  {\doibase 10.1093/ptep/pty120} {\bibfield  {journal} {\bibinfo  {journal}
  {PTEP}\ }\textbf {\bibinfo {volume} {2018}},\ \bibinfo {pages} {123E01}
  (\bibinfo {year} {2018})},\ \Eprint {http://arxiv.org/abs/1805.03946}
  {arXiv:1805.03946 [astro-ph.CO]} \BibitemShut {NoStop}%
\bibitem [{\citenamefont {Musco}(2019)}]{Musco:2018rwt}%
  \BibitemOpen
  \bibfield  {author} {\bibinfo {author} {\bibfnamefont {I.}~\bibnamefont
  {Musco}},\ }\href {\doibase 10.1103/PhysRevD.100.123524} {\bibfield
  {journal} {\bibinfo  {journal} {Phys. Rev. D}\ }\textbf {\bibinfo {volume}
  {100}},\ \bibinfo {pages} {123524} (\bibinfo {year} {2019})},\ \Eprint
  {http://arxiv.org/abs/1809.02127} {arXiv:1809.02127 [gr-qc]} \BibitemShut
  {NoStop}%
\bibitem [{\citenamefont {Kuhnel}\ \emph {et~al.}(2016)\citenamefont {Kuhnel},
  \citenamefont {Rampf},\ and\ \citenamefont {Sandstad}}]{Kuhnel:2015vtw}%
  \BibitemOpen
  \bibfield  {author} {\bibinfo {author} {\bibfnamefont {F.}~\bibnamefont
  {Kuhnel}}, \bibinfo {author} {\bibfnamefont {C.}~\bibnamefont {Rampf}}, \
  and\ \bibinfo {author} {\bibfnamefont {M.}~\bibnamefont {Sandstad}},\ }\href
  {\doibase 10.1140/epjc/s10052-016-3945-8} {\bibfield  {journal} {\bibinfo
  {journal} {Eur.~Phys.~J.}\ }\textbf {\bibinfo {volume} {C76}},\ \bibinfo
  {pages} {93} (\bibinfo {year} {2016})},\ \Eprint
  {http://arxiv.org/abs/1512.00488} {arXiv:1512.00488 [astro-ph.CO]}
  \BibitemShut {NoStop}%
\bibitem [{\citenamefont {Aghanim}\ \emph {et~al.}(2018)\citenamefont {Aghanim}
  \emph {et~al.}}]{Aghanim:2018eyx}%
  \BibitemOpen
  \bibfield  {author} {\bibinfo {author} {\bibfnamefont {N.}~\bibnamefont
  {Aghanim}} \emph {et~al.} (\bibinfo {collaboration} {Planck}),\ }\href@noop
  {} {\  (\bibinfo {year} {2018})},\ \Eprint {http://arxiv.org/abs/1807.06209}
  {arXiv:1807.06209 [astro-ph.CO]} \BibitemShut {NoStop}%
\bibitem [{\citenamefont {Ezquiaga}\ \emph {et~al.}(2018)\citenamefont
  {Ezquiaga}, \citenamefont {Garc{\'i}a-Bellido},\ and\ \citenamefont
  {Ruiz~Morales}}]{Ezquiaga:2017fvi}%
  \BibitemOpen
  \bibfield  {author} {\bibinfo {author} {\bibfnamefont {J.~M.}\ \bibnamefont
  {Ezquiaga}}, \bibinfo {author} {\bibfnamefont {J.}~\bibnamefont
  {Garc{\'i}a-Bellido}}, \ and\ \bibinfo {author} {\bibfnamefont
  {E.}~\bibnamefont {Ruiz~Morales}},\ }\href {\doibase
  10.1016/j.physletb.2017.11.039} {\bibfield  {journal} {\bibinfo  {journal}
  {Phys. Lett.}\ }\textbf {\bibinfo {volume} {B776}},\ \bibinfo {pages} {345}
  (\bibinfo {year} {2018})},\ \Eprint {http://arxiv.org/abs/1705.04861}
  {arXiv:1705.04861 [astro-ph.CO]} \BibitemShut {NoStop}%
\bibitem [{\citenamefont {Clesse}\ and\ \citenamefont
  {Garc{\'i}a-Bellido}(2015)}]{Clesse:2015wea}%
  \BibitemOpen
  \bibfield  {author} {\bibinfo {author} {\bibfnamefont {S.}~\bibnamefont
  {Clesse}}\ and\ \bibinfo {author} {\bibfnamefont {J.}~\bibnamefont
  {Garc{\'i}a-Bellido}},\ }\href {\doibase 10.1103/PhysRevD.92.023524}
  {\bibfield  {journal} {\bibinfo  {journal} {Phys.~Rev.}\ }\textbf {\bibinfo
  {volume} {D92}},\ \bibinfo {pages} {023524} (\bibinfo {year} {2015})},\
  \Eprint {http://arxiv.org/abs/1501.07565} {arXiv:1501.07565 [astro-ph.CO]}
  \BibitemShut {NoStop}%
\bibitem [{\citenamefont {Kalaja}\ \emph {et~al.}(2019)\citenamefont {Kalaja},
  \citenamefont {Bellomo}, \citenamefont {Bartolo}, \citenamefont {Bertacca},
  \citenamefont {Matarrese}, \citenamefont {Musco}, \citenamefont
  {Raccanelli},\ and\ \citenamefont {Verde}}]{Kalaja:2019uju}%
  \BibitemOpen
  \bibfield  {author} {\bibinfo {author} {\bibfnamefont {A.}~\bibnamefont
  {Kalaja}}, \bibinfo {author} {\bibfnamefont {N.}~\bibnamefont {Bellomo}},
  \bibinfo {author} {\bibfnamefont {N.}~\bibnamefont {Bartolo}}, \bibinfo
  {author} {\bibfnamefont {D.}~\bibnamefont {Bertacca}}, \bibinfo {author}
  {\bibfnamefont {S.}~\bibnamefont {Matarrese}}, \bibinfo {author}
  {\bibfnamefont {I.}~\bibnamefont {Musco}}, \bibinfo {author} {\bibfnamefont
  {A.}~\bibnamefont {Raccanelli}}, \ and\ \bibinfo {author} {\bibfnamefont
  {L.}~\bibnamefont {Verde}},\ }\href {\doibase 10.1088/1475-7516/2019/10/031}
  {\bibfield  {journal} {\bibinfo  {journal} {JCAP}\ }\textbf {\bibinfo
  {volume} {1910}},\ \bibinfo {pages} {031} (\bibinfo {year} {2019})},\ \Eprint
  {http://arxiv.org/abs/1908.03596} {arXiv:1908.03596 [astro-ph.CO]}
  \BibitemShut {NoStop}%
\bibitem [{\citenamefont {Jedamzik}(1997)}]{Jedamzik:1996mr}%
  \BibitemOpen
  \bibfield  {author} {\bibinfo {author} {\bibfnamefont {K.}~\bibnamefont
  {Jedamzik}},\ }\href {\doibase 10.1103/PhysRevD.55.5871} {\bibfield
  {journal} {\bibinfo  {journal} {Phys.~Rev.}\ }\textbf {\bibinfo {volume}
  {D55}},\ \bibinfo {pages} {R5871} (\bibinfo {year} {1997})}\BibitemShut
  {NoStop}%
\bibitem [{\citenamefont {Li}\ \emph {et~al.}(2017)\citenamefont {Li} \emph
  {et~al.}}]{Li:2016utv}%
  \BibitemOpen
  \bibfield  {author} {\bibinfo {author} {\bibfnamefont {T.~S.}\ \bibnamefont
  {Li}} \emph {et~al.} (\bibinfo {collaboration} {DES}),\ }\href {\doibase
  10.3847/1538-4357/aa6113} {\bibfield  {journal} {\bibinfo  {journal}
  {Astrophys. J.}\ }\textbf {\bibinfo {volume} {838}},\ \bibinfo {pages} {8}
  (\bibinfo {year} {2017})},\ \Eprint {http://arxiv.org/abs/1611.05052}
  {arXiv:1611.05052 [astro-ph.GA]} \BibitemShut {NoStop}%
\bibitem [{\citenamefont {Gaggero}\ \emph {et~al.}(2017)\citenamefont
  {Gaggero}, \citenamefont {Bertone}, \citenamefont {Calore}, \citenamefont
  {Connors}, \citenamefont {Lovell}, \citenamefont {Markoff},\ and\
  \citenamefont {Storm}}]{Gaggero:2016dpq}%
  \BibitemOpen
  \bibfield  {author} {\bibinfo {author} {\bibfnamefont {D.}~\bibnamefont
  {Gaggero}}, \bibinfo {author} {\bibfnamefont {G.}~\bibnamefont {Bertone}},
  \bibinfo {author} {\bibfnamefont {F.}~\bibnamefont {Calore}}, \bibinfo
  {author} {\bibfnamefont {R.~M.~T.}\ \bibnamefont {Connors}}, \bibinfo
  {author} {\bibfnamefont {M.}~\bibnamefont {Lovell}}, \bibinfo {author}
  {\bibfnamefont {S.}~\bibnamefont {Markoff}}, \ and\ \bibinfo {author}
  {\bibfnamefont {E.}~\bibnamefont {Storm}},\ }\href {\doibase
  10.1103/PhysRevLett.118.241101} {\bibfield  {journal} {\bibinfo  {journal}
  {Phys. Rev. Lett.}\ }\textbf {\bibinfo {volume} {118}},\ \bibinfo {pages}
  {241101} (\bibinfo {year} {2017})},\ \Eprint
  {http://arxiv.org/abs/1612.00457} {arXiv:1612.00457 [astro-ph.HE]}
  \BibitemShut {NoStop}%
\bibitem [{\citenamefont {Quinn}\ \emph {et~al.}(2009)\citenamefont {Quinn},
  \citenamefont {Wilkinson}, \citenamefont {Irwin}, \citenamefont {Marshall},
  \citenamefont {Koch},\ and\ \citenamefont {Belokurov}}]{Quinn:2009zg}%
  \BibitemOpen
  \bibfield  {author} {\bibinfo {author} {\bibfnamefont {D.~P.}\ \bibnamefont
  {Quinn}}, \bibinfo {author} {\bibfnamefont {M.~I.}\ \bibnamefont
  {Wilkinson}}, \bibinfo {author} {\bibfnamefont {M.~J.}\ \bibnamefont
  {Irwin}}, \bibinfo {author} {\bibfnamefont {J.}~\bibnamefont {Marshall}},
  \bibinfo {author} {\bibfnamefont {A.}~\bibnamefont {Koch}}, \ and\ \bibinfo
  {author} {\bibfnamefont {V.}~\bibnamefont {Belokurov}},\ }\href {\doibase
  10.1111/j.1745-3933.2009.00652.x} {\bibfield  {journal} {\bibinfo  {journal}
  {Mon. Not. Roy. Astron. Soc.}\ }\textbf {\bibinfo {volume} {396}},\ \bibinfo
  {pages} {11} (\bibinfo {year} {2009})},\ \Eprint
  {http://arxiv.org/abs/0903.1644} {arXiv:0903.1644 [astro-ph.GA]} \BibitemShut
  {NoStop}%
\bibitem [{\citenamefont {Ali-Ha{\"\i}moud}\ and\ \citenamefont
  {Kamionkowski}(2017)}]{Ali-Haimoud:2016mbv}%
  \BibitemOpen
  \bibfield  {author} {\bibinfo {author} {\bibfnamefont {Y.}~\bibnamefont
  {Ali-Ha{\"\i}moud}}\ and\ \bibinfo {author} {\bibfnamefont {M.}~\bibnamefont
  {Kamionkowski}},\ }\href {\doibase 10.1103/PhysRevD.95.043534} {\bibfield
  {journal} {\bibinfo  {journal} {Phys. Rev.}\ }\textbf {\bibinfo {volume}
  {D95}},\ \bibinfo {pages} {043534} (\bibinfo {year} {2017})},\ \Eprint
  {http://arxiv.org/abs/1612.05644} {arXiv:1612.05644 [astro-ph.CO]}
  \BibitemShut {NoStop}%
\bibitem [{\citenamefont {Carr}\ \emph {et~al.}(2016)\citenamefont {Carr},
  \citenamefont {Kuhnel},\ and\ \citenamefont {Sandstad}}]{Carr:2016drx}%
  \BibitemOpen
  \bibfield  {author} {\bibinfo {author} {\bibfnamefont {B.}~\bibnamefont
  {Carr}}, \bibinfo {author} {\bibfnamefont {F.}~\bibnamefont {Kuhnel}}, \ and\
  \bibinfo {author} {\bibfnamefont {M.}~\bibnamefont {Sandstad}},\ }\href
  {\doibase 10.1103/PhysRevD.94.083504} {\bibfield  {journal} {\bibinfo
  {journal} {Phys.~Rev.}\ }\textbf {\bibinfo {volume} {D94}},\ \bibinfo {pages}
  {083504} (\bibinfo {year} {2016})}\BibitemShut {NoStop}%
\bibitem [{\citenamefont {Carr}\ \emph {et~al.}(2017)\citenamefont {Carr},
  \citenamefont {Raidal}, \citenamefont {Tenkanen}, \citenamefont {Vaskonen},\
  and\ \citenamefont {Veerm{\"a}e}}]{2017PhRvD..96b3514C}%
  \BibitemOpen
  \bibfield  {author} {\bibinfo {author} {\bibfnamefont {B.}~\bibnamefont
  {Carr}}, \bibinfo {author} {\bibfnamefont {M.}~\bibnamefont {Raidal}},
  \bibinfo {author} {\bibfnamefont {T.}~\bibnamefont {Tenkanen}}, \bibinfo
  {author} {\bibfnamefont {V.}~\bibnamefont {Vaskonen}}, \ and\ \bibinfo
  {author} {\bibfnamefont {H.}~\bibnamefont {Veerm{\"a}e}},\ }\href {\doibase
  10.1103/PhysRevD.96.023514} {\bibfield  {journal} {\bibinfo  {journal} {Phys.
  Rev.}\ }\textbf {\bibinfo {volume} {D96}},\ \bibinfo {pages} {023514}
  (\bibinfo {year} {2017})},\ \Eprint {http://arxiv.org/abs/1705.05567}
  {arXiv:1705.05567 [astro-ph.CO]} \BibitemShut {NoStop}%
\bibitem [{\citenamefont {Poulin}\ \emph {et~al.}(2017)\citenamefont {Poulin},
  \citenamefont {Serpico}, \citenamefont {Calore}, \citenamefont {Clesse},\
  and\ \citenamefont {Kohri}}]{Poulin:2017bwe}%
  \BibitemOpen
  \bibfield  {author} {\bibinfo {author} {\bibfnamefont {V.}~\bibnamefont
  {Poulin}}, \bibinfo {author} {\bibfnamefont {P.~D.}\ \bibnamefont {Serpico}},
  \bibinfo {author} {\bibfnamefont {F.}~\bibnamefont {Calore}}, \bibinfo
  {author} {\bibfnamefont {S.}~\bibnamefont {Clesse}}, \ and\ \bibinfo {author}
  {\bibfnamefont {K.}~\bibnamefont {Kohri}},\ }\href {\doibase
  10.1103/PhysRevD.96.083524} {\bibfield  {journal} {\bibinfo  {journal} {Phys.
  Rev.}\ }\textbf {\bibinfo {volume} {D96}},\ \bibinfo {pages} {083524}
  (\bibinfo {year} {2017})},\ \Eprint {http://arxiv.org/abs/1707.04206}
  {arXiv:1707.04206 [astro-ph.CO]} \BibitemShut {NoStop}%
\bibitem [{\citenamefont {Byrnes}\ \emph {et~al.}(2019)\citenamefont {Byrnes},
  \citenamefont {Cole},\ and\ \citenamefont {Patil}}]{Byrnes:2018txb}%
  \BibitemOpen
  \bibfield  {author} {\bibinfo {author} {\bibfnamefont {C.~T.}\ \bibnamefont
  {Byrnes}}, \bibinfo {author} {\bibfnamefont {P.~S.}\ \bibnamefont {Cole}}, \
  and\ \bibinfo {author} {\bibfnamefont {S.~P.}\ \bibnamefont {Patil}},\ }\href
  {\doibase 10.1088/1475-7516/2019/06/028} {\bibfield  {journal} {\bibinfo
  {journal} {JCAP}\ }\textbf {\bibinfo {volume} {1906}},\ \bibinfo {pages}
  {028} (\bibinfo {year} {2019})},\ \Eprint {http://arxiv.org/abs/1811.11158}
  {arXiv:1811.11158 [astro-ph.CO]} \BibitemShut {NoStop}%
\bibitem [{\citenamefont {Nakama}\ \emph {et~al.}(2018)\citenamefont {Nakama},
  \citenamefont {Carr},\ and\ \citenamefont {Silk}}]{Nakama:2017xvq}%
  \BibitemOpen
  \bibfield  {author} {\bibinfo {author} {\bibfnamefont {T.}~\bibnamefont
  {Nakama}}, \bibinfo {author} {\bibfnamefont {B.}~\bibnamefont {Carr}}, \ and\
  \bibinfo {author} {\bibfnamefont {J.}~\bibnamefont {Silk}},\ }\href {\doibase
  10.1103/PhysRevD.97.043525} {\bibfield  {journal} {\bibinfo  {journal} {Phys.
  Rev.}\ }\textbf {\bibinfo {volume} {D97}},\ \bibinfo {pages} {043525}
  (\bibinfo {year} {2018})},\ \Eprint {http://arxiv.org/abs/1710.06945}
  {arXiv:1710.06945 [astro-ph.CO]} \BibitemShut {NoStop}%
\bibitem [{\citenamefont {Ezquiaga}\ and\ \citenamefont
  {Garc{\'i}a-Bellido}(2018)}]{Ezquiaga:2018gbw}%
  \BibitemOpen
  \bibfield  {author} {\bibinfo {author} {\bibfnamefont {J.~M.}\ \bibnamefont
  {Ezquiaga}}\ and\ \bibinfo {author} {\bibfnamefont {J.}~\bibnamefont
  {Garc{\'i}a-Bellido}},\ }\href {\doibase 10.1088/1475-7516/2018/08/018}
  {\bibfield  {journal} {\bibinfo  {journal} {JCAP}\ }\textbf {\bibinfo
  {volume} {1808}},\ \bibinfo {pages} {018} (\bibinfo {year} {2018})},\ \Eprint
  {http://arxiv.org/abs/1805.06731} {arXiv:1805.06731 [astro-ph.CO]}
  \BibitemShut {NoStop}%
\bibitem [{\citenamefont {Ezquiaga}\ \emph {et~al.}(2020)\citenamefont
  {Ezquiaga}, \citenamefont {Garc\'\i{}a-Bellido},\ and\ \citenamefont
  {Vennin}}]{Ezquiaga:2019ftu}%
  \BibitemOpen
  \bibfield  {author} {\bibinfo {author} {\bibfnamefont {J.~M.}\ \bibnamefont
  {Ezquiaga}}, \bibinfo {author} {\bibfnamefont {J.}~\bibnamefont
  {Garc\'\i{}a-Bellido}}, \ and\ \bibinfo {author} {\bibfnamefont
  {V.}~\bibnamefont {Vennin}},\ }\href {\doibase 10.1088/1475-7516/2020/03/029}
  {\bibfield  {journal} {\bibinfo  {journal} {JCAP}\ }\textbf {\bibinfo
  {volume} {03}},\ \bibinfo {pages} {029} (\bibinfo {year} {2020})},\ \Eprint
  {http://arxiv.org/abs/1912.05399} {arXiv:1912.05399 [astro-ph.CO]}
  \BibitemShut {NoStop}%
\bibitem [{\citenamefont {Tisserand}\ \emph {et~al.}(2007)\citenamefont
  {Tisserand} \emph {et~al.}}]{Tisserand:2006zx}%
  \BibitemOpen
  \bibfield  {author} {\bibinfo {author} {\bibfnamefont {P.}~\bibnamefont
  {Tisserand}} \emph {et~al.} (\bibinfo {collaboration} {EROS-2}),\ }\href
  {\doibase 10.1051/0004-6361:20066017} {\bibfield  {journal} {\bibinfo
  {journal} {Astron. astrophys.}\ }\textbf {\bibinfo {volume} {469}},\ \bibinfo
  {pages} {387} (\bibinfo {year} {2007})},\ \Eprint
  {http://arxiv.org/abs/astro-ph/0607207} {arXiv:astro-ph/0607207} \BibitemShut
  {NoStop}%
\bibitem [{\citenamefont {Wyrzykowski}\ \emph {et~al.}(2009)\citenamefont
  {Wyrzykowski} \emph {et~al.}}]{Wyrzykowski:2009ep}%
  \BibitemOpen
  \bibfield  {author} {\bibinfo {author} {\bibfnamefont {L.}~\bibnamefont
  {Wyrzykowski}} \emph {et~al.},\ }\href@noop {} {\  (\bibinfo {year}
  {2009})},\ \Eprint {http://arxiv.org/abs/0905.2044} {arXiv:0905.2044
  [astro-ph.GA]} \BibitemShut {NoStop}%
\bibitem [{\citenamefont {Green}(2016)}]{Green:2016xgy}%
  \BibitemOpen
  \bibfield  {author} {\bibinfo {author} {\bibfnamefont {A.~M.}\ \bibnamefont
  {Green}},\ }\href {\doibase 10.1103/PhysRevD.94.063530} {\bibfield  {journal}
  {\bibinfo  {journal} {Phys. Rev.}\ }\textbf {\bibinfo {volume} {D94}},\
  \bibinfo {pages} {063530} (\bibinfo {year} {2016})},\ \Eprint
  {http://arxiv.org/abs/1609.01143} {arXiv:1609.01143 [astro-ph.CO]}
  \BibitemShut {NoStop}%
\bibitem [{\citenamefont {Calcino}\ \emph {et~al.}(2018)\citenamefont
  {Calcino}, \citenamefont {Garc{\'i}a-Bellido},\ and\ \citenamefont
  {Davis}}]{Calcino:2018mwh}%
  \BibitemOpen
  \bibfield  {author} {\bibinfo {author} {\bibfnamefont {J.}~\bibnamefont
  {Calcino}}, \bibinfo {author} {\bibfnamefont {J.}~\bibnamefont
  {Garc{\'i}a-Bellido}}, \ and\ \bibinfo {author} {\bibfnamefont {T.~M.}\
  \bibnamefont {Davis}},\ }\href {\doibase 10.1093/mnras/sty1368} {\bibfield
  {journal} {\bibinfo  {journal} {Mon. Not. Roy. Astron. Soc.}\ }\textbf
  {\bibinfo {volume} {479}},\ \bibinfo {pages} {2889} (\bibinfo {year}
  {2018})},\ \Eprint {http://arxiv.org/abs/1803.09205} {arXiv:1803.09205
  [astro-ph.CO]} \BibitemShut {NoStop}%
\bibitem [{\citenamefont {Brandt}(2016)}]{Brandt:2016aco}%
  \BibitemOpen
  \bibfield  {author} {\bibinfo {author} {\bibfnamefont {T.~D.}\ \bibnamefont
  {Brandt}},\ }\href {\doibase 10.3847/2041-8205/824/2/L31} {\bibfield
  {journal} {\bibinfo  {journal} {Astrophys.~J.}\ }\textbf {\bibinfo {volume}
  {824}},\ \bibinfo {pages} {L31} (\bibinfo {year} {2016})},\ \Eprint
  {http://arxiv.org/abs/1605.03665} {arXiv:1605.03665 [astro-ph.GA]}
  \BibitemShut {NoStop}%
\bibitem [{\citenamefont {Niikura}\ \emph {et~al.}(2017)\citenamefont {Niikura}
  \emph {et~al.}}]{Niikura:2017zjd}%
  \BibitemOpen
  \bibfield  {author} {\bibinfo {author} {\bibfnamefont {H.}~\bibnamefont
  {Niikura}} \emph {et~al.},\ }\href {\doibase 10.1038/s41550-019-0723-1} {\
  (\bibinfo {year} {2017}),\ 10.1038/s41550-019-0723-1},\ \Eprint
  {http://arxiv.org/abs/1701.02151} {arXiv:1701.02151 [astro-ph.CO]}
  \BibitemShut {NoStop}%
\bibitem [{\citenamefont {{Mr{\'o}z}}\ \emph {et~al.}(2017)\citenamefont
  {{Mr{\'o}z}}, \citenamefont {{Udalski}}, \citenamefont {{Skowron}},
  \citenamefont {{Poleski}}, \citenamefont {{Koz{\l}owski}}, \citenamefont
  {{Szyma{\'n}ski}}, \citenamefont {{Soszy{\'n}ski}}, \citenamefont
  {{Wyrzykowski}}, \citenamefont {{Pietrukowicz}}, \citenamefont {{Ulaczyk}},
  \citenamefont {{Skowron}},\ and\ \citenamefont
  {{Pawlak}}}]{2017Natur.548..183M}%
  \BibitemOpen
  \bibfield  {author} {\bibinfo {author} {\bibfnamefont {P.}~\bibnamefont
  {{Mr{\'o}z}}}, \bibinfo {author} {\bibfnamefont {A.}~\bibnamefont
  {{Udalski}}}, \bibinfo {author} {\bibfnamefont {J.}~\bibnamefont
  {{Skowron}}}, \bibinfo {author} {\bibfnamefont {R.}~\bibnamefont
  {{Poleski}}}, \bibinfo {author} {\bibfnamefont {S.}~\bibnamefont
  {{Koz{\l}owski}}}, \bibinfo {author} {\bibfnamefont {M.~K.}\ \bibnamefont
  {{Szyma{\'n}ski}}}, \bibinfo {author} {\bibfnamefont {I.}~\bibnamefont
  {{Soszy{\'n}ski}}}, \bibinfo {author} {\bibfnamefont {{\L}.}~\bibnamefont
  {{Wyrzykowski}}}, \bibinfo {author} {\bibfnamefont {P.}~\bibnamefont
  {{Pietrukowicz}}}, \bibinfo {author} {\bibfnamefont {K.}~\bibnamefont
  {{Ulaczyk}}}, \bibinfo {author} {\bibfnamefont {D.}~\bibnamefont
  {{Skowron}}}, \ and\ \bibinfo {author} {\bibfnamefont {M.}~\bibnamefont
  {{Pawlak}}},\ }\href {\doibase 10.1038/nature23276} {\bibfield  {journal}
  {\bibinfo  {journal} {\nat}\ }\textbf {\bibinfo {volume} {548}},\ \bibinfo
  {pages} {183} (\bibinfo {year} {2017})},\ \Eprint
  {http://arxiv.org/abs/1707.07634} {arXiv:1707.07634 [astro-ph.EP]}
  \BibitemShut {NoStop}%
\bibitem [{\citenamefont {{van Elteren}}\ \emph {et~al.}(2019)\citenamefont
  {{van Elteren}}, \citenamefont {{Portegies Zwart}}, \citenamefont
  {{Pelupessy}}, \citenamefont {{Cai}},\ and\ \citenamefont
  {{McMillan}}}]{2019A&A...624A.120V}%
  \BibitemOpen
  \bibfield  {author} {\bibinfo {author} {\bibfnamefont {A.}~\bibnamefont {{van
  Elteren}}}, \bibinfo {author} {\bibfnamefont {S.}~\bibnamefont {{Portegies
  Zwart}}}, \bibinfo {author} {\bibfnamefont {I.}~\bibnamefont {{Pelupessy}}},
  \bibinfo {author} {\bibfnamefont {M.~X.}\ \bibnamefont {{Cai}}}, \ and\
  \bibinfo {author} {\bibfnamefont {S.~L.~W.}\ \bibnamefont {{McMillan}}},\
  }\href {\doibase 10.1051/0004-6361/201834641} {\bibfield  {journal} {\bibinfo
   {journal} {A \& A}\ }\textbf {\bibinfo {volume} {624}},\ \bibinfo {eid}
  {A120} (\bibinfo {year} {2019})},\ \Eprint {http://arxiv.org/abs/1902.04652}
  {arXiv:1902.04652 [astro-ph.SR]} \BibitemShut {NoStop}%
\bibitem [{\citenamefont {{Hawkins}}(1993)}]{1993Natur.366..242H}%
  \BibitemOpen
  \bibfield  {author} {\bibinfo {author} {\bibfnamefont {M.~R.~S.}\
  \bibnamefont {{Hawkins}}},\ }\href {\doibase 10.1038/366242a0} {\bibfield
  {journal} {\bibinfo  {journal} {Nature}\ }\textbf {\bibinfo {volume} {366}},\
  \bibinfo {pages} {242} (\bibinfo {year} {1993})}\BibitemShut {NoStop}%
\bibitem [{\citenamefont {Hawkins}(2006)}]{Hawkins:2006xj}%
  \BibitemOpen
  \bibfield  {author} {\bibinfo {author} {\bibfnamefont {M.~R.~S.}\
  \bibnamefont {Hawkins}},\ }\href {\doibase 10.1051/0004-6361:20066283}
  {\bibfield  {journal} {\bibinfo  {journal} {Astron. Astrophys.}\ } (\bibinfo
  {year} {2006}),\ 10.1051/0004-6361:20066283},\ \bibinfo {note} {[Astron.
  Astrophys.462,581(2007)]},\ \Eprint {http://arxiv.org/abs/astro-ph/0611491}
  {arXiv:astro-ph/0611491 [astro-ph]} \BibitemShut {NoStop}%
\bibitem [{\citenamefont {Hawkins}(2020)}]{Hawkins:2020zie}%
  \BibitemOpen
  \bibfield  {author} {\bibinfo {author} {\bibfnamefont {M.~R.~S.}\
  \bibnamefont {Hawkins}},\ }\href {\doibase 10.1051/0004-6361/201936462}
  {\bibfield  {journal} {\bibinfo  {journal} {Astron. Astrophys.}\ }\textbf
  {\bibinfo {volume} {633}},\ \bibinfo {pages} {A107} (\bibinfo {year}
  {2020})},\ \Eprint {http://arxiv.org/abs/2001.07633} {arXiv:2001.07633
  [astro-ph.GA]} \BibitemShut {NoStop}%
\bibitem [{\citenamefont {Kashlinsky}(2016)}]{Kashlinsky:2016sdv}%
  \BibitemOpen
  \bibfield  {author} {\bibinfo {author} {\bibfnamefont {A.}~\bibnamefont
  {Kashlinsky}},\ }\href {\doibase 10.3847/2041-8205/823/2/L25} {\bibfield
  {journal} {\bibinfo  {journal} {Astrophys.~J.}\ }\textbf {\bibinfo {volume}
  {823}},\ \bibinfo {pages} {L25} (\bibinfo {year} {2016})},\ \Eprint
  {http://arxiv.org/abs/1605.04023} {arXiv:1605.04023 [astro-ph.CO]}
  \BibitemShut {NoStop}%
\bibitem [{\citenamefont {Boldrini}\ \emph {et~al.}(2020)\citenamefont
  {Boldrini}, \citenamefont {Miki}, \citenamefont {Wagner}, \citenamefont
  {Mohayaee}, \citenamefont {Silk},\ and\ \citenamefont
  {Arbey}}]{Boldrini:2019isx}%
  \BibitemOpen
  \bibfield  {author} {\bibinfo {author} {\bibfnamefont {P.}~\bibnamefont
  {Boldrini}}, \bibinfo {author} {\bibfnamefont {Y.}~\bibnamefont {Miki}},
  \bibinfo {author} {\bibfnamefont {A.~Y.}\ \bibnamefont {Wagner}}, \bibinfo
  {author} {\bibfnamefont {R.}~\bibnamefont {Mohayaee}}, \bibinfo {author}
  {\bibfnamefont {J.}~\bibnamefont {Silk}}, \ and\ \bibinfo {author}
  {\bibfnamefont {A.}~\bibnamefont {Arbey}},\ }\href {\doibase
  10.1093/mnras/staa150} {\bibfield  {journal} {\bibinfo  {journal} {Mon. Not.
  Roy. Astron. Soc.}\ }\textbf {\bibinfo {volume} {492}},\ \bibinfo {pages}
  {5218} (\bibinfo {year} {2020})},\ \Eprint {http://arxiv.org/abs/1909.07395}
  {arXiv:1909.07395 [astro-ph.CO]} \BibitemShut {NoStop}%
\bibitem [{\citenamefont {Abbott}\ \emph
  {et~al.}(2019{\natexlab{a}})\citenamefont {Abbott} \emph
  {et~al.}}]{LIGOScientific:2018jsj}%
  \BibitemOpen
  \bibfield  {author} {\bibinfo {author} {\bibfnamefont {B.}~\bibnamefont
  {Abbott}} \emph {et~al.} (\bibinfo {collaboration} {LIGO Scientific,
  Virgo}),\ }\href {\doibase 10.3847/2041-8213/ab3800} {\bibfield  {journal}
  {\bibinfo  {journal} {Astrophys. J. Lett.}\ }\textbf {\bibinfo {volume}
  {882}},\ \bibinfo {pages} {L24} (\bibinfo {year} {2019}{\natexlab{a}})},\
  \Eprint {http://arxiv.org/abs/1811.12940} {arXiv:1811.12940 [astro-ph.HE]}
  \BibitemShut {NoStop}%
\bibitem [{\citenamefont {Abbott}\ \emph
  {et~al.}(2020{\natexlab{a}})\citenamefont {Abbott} \emph
  {et~al.}}]{Abbott:2020uma}%
  \BibitemOpen
  \bibfield  {author} {\bibinfo {author} {\bibfnamefont {B.~P.}\ \bibnamefont
  {Abbott}} \emph {et~al.} (\bibinfo {collaboration} {LIGO Scientific,
  Virgo}),\ }\href@noop {} {\  (\bibinfo {year} {2020}{\natexlab{a}})},\
  \Eprint {http://arxiv.org/abs/2001.01761} {arXiv:2001.01761 [astro-ph.HE]}
  \BibitemShut {NoStop}%
\bibitem [{\citenamefont {Abbott}\ \emph
  {et~al.}(2020{\natexlab{b}})\citenamefont {Abbott} \emph
  {et~al.}}]{Abbott:2020khf}%
  \BibitemOpen
  \bibfield  {author} {\bibinfo {author} {\bibfnamefont {R.}~\bibnamefont
  {Abbott}} \emph {et~al.} (\bibinfo {collaboration} {LIGO Scientific,
  Virgo}),\ }\href {\doibase 10.3847/2041-8213/ab960f} {\bibfield  {journal}
  {\bibinfo  {journal} {Astrophys. J. Lett.}\ }\textbf {\bibinfo {volume}
  {896}},\ \bibinfo {pages} {L44} (\bibinfo {year} {2020}{\natexlab{b}})},\
  \Eprint {http://arxiv.org/abs/2006.12611} {arXiv:2006.12611 [astro-ph.HE]}
  \BibitemShut {NoStop}%
\bibitem [{\citenamefont {Jedamzik}(2020)}]{Jedamzik:2020omx}%
  \BibitemOpen
  \bibfield  {author} {\bibinfo {author} {\bibfnamefont {K.}~\bibnamefont
  {Jedamzik}},\ }\href@noop {} {\  (\bibinfo {year} {2020})},\ \Eprint
  {http://arxiv.org/abs/2007.03565} {arXiv:2007.03565 [astro-ph.CO]}
  \BibitemShut {NoStop}%
\bibitem [{\citenamefont {Abbott}\ \emph
  {et~al.}(2020{\natexlab{c}})\citenamefont {Abbott} \emph
  {et~al.}}]{Abbott:2020tfl}%
  \BibitemOpen
  \bibfield  {author} {\bibinfo {author} {\bibfnamefont {R.}~\bibnamefont
  {Abbott}} \emph {et~al.} (\bibinfo {collaboration} {LIGO Scientific,
  Virgo}),\ }\href {\doibase 10.1103/PhysRevLett.125.101102} {\bibfield
  {journal} {\bibinfo  {journal} {Phys. Rev. Lett.}\ }\textbf {\bibinfo
  {volume} {125}},\ \bibinfo {pages} {101102} (\bibinfo {year}
  {2020}{\natexlab{c}})},\ \Eprint {http://arxiv.org/abs/2009.01075}
  {arXiv:2009.01075 [gr-qc]} \BibitemShut {NoStop}%
\bibitem [{\citenamefont {Abbott}\ \emph
  {et~al.}(2020{\natexlab{d}})\citenamefont {Abbott} \emph
  {et~al.}}]{Abbott:2020mjq}%
  \BibitemOpen
  \bibfield  {author} {\bibinfo {author} {\bibfnamefont {R.}~\bibnamefont
  {Abbott}} \emph {et~al.} (\bibinfo {collaboration} {LIGO Scientific,
  Virgo}),\ }\href {\doibase 10.3847/2041-8213/aba493} {\bibfield  {journal}
  {\bibinfo  {journal} {Astrophys. J. Lett.}\ }\textbf {\bibinfo {volume}
  {900}},\ \bibinfo {pages} {L13} (\bibinfo {year} {2020}{\natexlab{d}})},\
  \Eprint {http://arxiv.org/abs/2009.01190} {arXiv:2009.01190 [astro-ph.HE]}
  \BibitemShut {NoStop}%
\bibitem [{\citenamefont {Press}\ and\ \citenamefont
  {Schechter}(1974)}]{Press:1973iz}%
  \BibitemOpen
  \bibfield  {author} {\bibinfo {author} {\bibfnamefont {W.~H.}\ \bibnamefont
  {Press}}\ and\ \bibinfo {author} {\bibfnamefont {P.}~\bibnamefont
  {Schechter}},\ }\href {\doibase 10.1086/152650} {\bibfield  {journal}
  {\bibinfo  {journal} {Astrophys. J.}\ }\textbf {\bibinfo {volume} {187}},\
  \bibinfo {pages} {425} (\bibinfo {year} {1974})}\BibitemShut {NoStop}%
\bibitem [{\citenamefont {Kruijssen}\ and\ \citenamefont
  {Lutzgendorf}(2013)}]{Kruijssen:2013cna}%
  \BibitemOpen
  \bibfield  {author} {\bibinfo {author} {\bibfnamefont {J.~M.~D.}\
  \bibnamefont {Kruijssen}}\ and\ \bibinfo {author} {\bibfnamefont
  {N.}~\bibnamefont {Lutzgendorf}},\ }\href {\doibase 10.1093/mnrasl/slt073}
  {\bibfield  {journal} {\bibinfo  {journal} {Mon. Not. Roy. Astron. Soc.}\
  }\textbf {\bibinfo {volume} {434}},\ \bibinfo {pages} {41} (\bibinfo {year}
  {2013})},\ \Eprint {http://arxiv.org/abs/1306.0561} {arXiv:1306.0561
  [astro-ph.CO]} \BibitemShut {NoStop}%
\bibitem [{\citenamefont {Kashlinsky}\ \emph {et~al.}(2019)\citenamefont
  {Kashlinsky} \emph {et~al.}}]{Ali-Haimoud:2019khd}%
  \BibitemOpen
  \bibfield  {author} {\bibinfo {author} {\bibfnamefont {A.}~\bibnamefont
  {Kashlinsky}} \emph {et~al.},\ }\href@noop {} {\  (\bibinfo {year} {2019})},\
  \Eprint {http://arxiv.org/abs/1903.04424} {arXiv:1903.04424 [astro-ph.CO]}
  \BibitemShut {NoStop}%
\bibitem [{\citenamefont {Kohri}\ and\ \citenamefont
  {Terada}(2020)}]{Kohri:2020qqd}%
  \BibitemOpen
  \bibfield  {author} {\bibinfo {author} {\bibfnamefont {K.}~\bibnamefont
  {Kohri}}\ and\ \bibinfo {author} {\bibfnamefont {T.}~\bibnamefont {Terada}},\
  }\href@noop {} {\  (\bibinfo {year} {2020})},\ \Eprint
  {http://arxiv.org/abs/2009.11853} {arXiv:2009.11853 [astro-ph.CO]}
  \BibitemShut {NoStop}%
\bibitem [{\citenamefont {De~Luca}\ \emph {et~al.}(2020)\citenamefont
  {De~Luca}, \citenamefont {Franciolini},\ and\ \citenamefont
  {Riotto}}]{DeLuca:2020agl}%
  \BibitemOpen
  \bibfield  {author} {\bibinfo {author} {\bibfnamefont {V.}~\bibnamefont
  {De~Luca}}, \bibinfo {author} {\bibfnamefont {G.}~\bibnamefont
  {Franciolini}}, \ and\ \bibinfo {author} {\bibfnamefont {A.}~\bibnamefont
  {Riotto}},\ }\href@noop {} {\  (\bibinfo {year} {2020})},\ \Eprint
  {http://arxiv.org/abs/2009.08268} {arXiv:2009.08268 [astro-ph.CO]}
  \BibitemShut {NoStop}%
\bibitem [{\citenamefont {Vaskonen}\ and\ \citenamefont
  {Veerm{\"a}e}(2020)}]{Vaskonen:2020lbd}%
  \BibitemOpen
  \bibfield  {author} {\bibinfo {author} {\bibfnamefont {V.}~\bibnamefont
  {Vaskonen}}\ and\ \bibinfo {author} {\bibfnamefont {H.}~\bibnamefont
  {Veerm{\"a}e}},\ }\href@noop {} {\  (\bibinfo {year} {2020})},\ \Eprint
  {http://arxiv.org/abs/2009.07832} {arXiv:2009.07832 [astro-ph.CO]}
  \BibitemShut {NoStop}%
\bibitem [{\citenamefont {Dom{\`e}nech}\ and\ \citenamefont
  {Pi}(2020)}]{Domenech:2020ers}%
  \BibitemOpen
  \bibfield  {author} {\bibinfo {author} {\bibfnamefont {G.}~\bibnamefont
  {Dom{\`e}nech}}\ and\ \bibinfo {author} {\bibfnamefont {S.}~\bibnamefont
  {Pi}},\ }\href@noop {} {\  (\bibinfo {year} {2020})},\ \Eprint
  {http://arxiv.org/abs/2010.03976} {arXiv:2010.03976 [astro-ph.CO]}
  \BibitemShut {NoStop}%
\bibitem [{\citenamefont {Bullock}\ and\ \citenamefont
  {Primack}(1997)}]{Bullock:1996at}%
  \BibitemOpen
  \bibfield  {author} {\bibinfo {author} {\bibfnamefont {J.~S.}\ \bibnamefont
  {Bullock}}\ and\ \bibinfo {author} {\bibfnamefont {J.~R.}\ \bibnamefont
  {Primack}},\ }\href {\doibase 10.1103/PhysRevD.55.7423} {\bibfield  {journal}
  {\bibinfo  {journal} {Phys. Rev.}\ }\textbf {\bibinfo {volume} {D55}},\
  \bibinfo {pages} {7423} (\bibinfo {year} {1997})},\ \Eprint
  {http://arxiv.org/abs/astro-ph/9611106} {arXiv:astro-ph/9611106 [astro-ph]}
  \BibitemShut {NoStop}%
\bibitem [{\citenamefont {Garc{\'i}a-Bellido}\ and\ \citenamefont
  {Ruiz~Morales}(2017)}]{Garcia-Bellido:2017mdw}%
  \BibitemOpen
  \bibfield  {author} {\bibinfo {author} {\bibfnamefont {J.}~\bibnamefont
  {Garc{\'i}a-Bellido}}\ and\ \bibinfo {author} {\bibfnamefont
  {E.}~\bibnamefont {Ruiz~Morales}},\ }\href {\doibase
  10.1016/j.dark.2017.09.007} {\bibfield  {journal} {\bibinfo  {journal} {Phys.
  Dark Univ.}\ }\textbf {\bibinfo {volume} {18}},\ \bibinfo {pages} {47}
  (\bibinfo {year} {2017})},\ \Eprint {http://arxiv.org/abs/1702.03901}
  {arXiv:1702.03901 [astro-ph.CO]} \BibitemShut {NoStop}%
\bibitem [{\citenamefont {Hawking}(1982)}]{Hawking:1982cz}%
  \BibitemOpen
  \bibfield  {author} {\bibinfo {author} {\bibfnamefont {S.~W.}\ \bibnamefont
  {Hawking}},\ }\href {\doibase 10.1016/0370-2693(82)90373-2} {\bibfield
  {journal} {\bibinfo  {journal} {Phys. Lett.}\ }\textbf {\bibinfo {volume}
  {115B}},\ \bibinfo {pages} {295} (\bibinfo {year} {1982})}\BibitemShut
  {NoStop}%
\bibitem [{\citenamefont {Carr}\ and\ \citenamefont
  {Silk}(2018)}]{Carr:2018rid}%
  \BibitemOpen
  \bibfield  {author} {\bibinfo {author} {\bibfnamefont {B.}~\bibnamefont
  {Carr}}\ and\ \bibinfo {author} {\bibfnamefont {J.}~\bibnamefont {Silk}},\
  }\href {\doibase 10.1093/mnras/sty1204} {\bibfield  {journal} {\bibinfo
  {journal} {Mon. Not. Roy. Astron. Soc.}\ }\textbf {\bibinfo {volume} {478}},\
  \bibinfo {pages} {3756} (\bibinfo {year} {2018})},\ \Eprint
  {http://arxiv.org/abs/1801.00672} {arXiv:1801.00672 [astro-ph.CO]}
  \BibitemShut {NoStop}%
\bibitem [{\citenamefont {Fleury}\ and\ \citenamefont
  {Garc{\'i}a-Bellido}(2019)}]{Fleury:2019xzr}%
  \BibitemOpen
  \bibfield  {author} {\bibinfo {author} {\bibfnamefont {P.}~\bibnamefont
  {Fleury}}\ and\ \bibinfo {author} {\bibfnamefont {J.}~\bibnamefont
  {Garc{\'i}a-Bellido}},\ }\href@noop {} {\  (\bibinfo {year} {2019})},\
  \Eprint {http://arxiv.org/abs/1907.05163} {arXiv:1907.05163 [astro-ph.CO]}
  \BibitemShut {NoStop}%
\bibitem [{\citenamefont {Zumalacarregui}\ and\ \citenamefont
  {Seljak}(2018)}]{Zumalacarregui:2017qqd}%
  \BibitemOpen
  \bibfield  {author} {\bibinfo {author} {\bibfnamefont {M.}~\bibnamefont
  {Zumalacarregui}}\ and\ \bibinfo {author} {\bibfnamefont {U.}~\bibnamefont
  {Seljak}},\ }\href {\doibase 10.1103/PhysRevLett.121.141101} {\bibfield
  {journal} {\bibinfo  {journal} {Phys. Rev. Lett.}\ }\textbf {\bibinfo
  {volume} {121}},\ \bibinfo {pages} {141101} (\bibinfo {year} {2018})},\
  \Eprint {http://arxiv.org/abs/1712.02240} {arXiv:1712.02240 [astro-ph.CO]}
  \BibitemShut {NoStop}%
\bibitem [{\citenamefont {Garc{\'i}a-Bellido}\ \emph
  {et~al.}(2018)\citenamefont {Garc{\'i}a-Bellido}, \citenamefont {Clesse},\
  and\ \citenamefont {Fleury}}]{Garcia-Bellido:2017imq}%
  \BibitemOpen
  \bibfield  {author} {\bibinfo {author} {\bibfnamefont {J.}~\bibnamefont
  {Garc{\'i}a-Bellido}}, \bibinfo {author} {\bibfnamefont {S.}~\bibnamefont
  {Clesse}}, \ and\ \bibinfo {author} {\bibfnamefont {P.}~\bibnamefont
  {Fleury}},\ }\href {\doibase 10.1016/j.dark.2018.04.005} {\bibfield
  {journal} {\bibinfo  {journal} {Phys. Dark Univ.}\ }\textbf {\bibinfo
  {volume} {20}},\ \bibinfo {pages} {95} (\bibinfo {year} {2018})},\ \Eprint
  {http://arxiv.org/abs/1712.06574} {arXiv:1712.06574 [astro-ph.CO]}
  \BibitemShut {NoStop}%
\bibitem [{\citenamefont {Clesse}\ and\ \citenamefont
  {Garc{\'i}a-Bellido}(2017)}]{Clesse:2016vqa}%
  \BibitemOpen
  \bibfield  {author} {\bibinfo {author} {\bibfnamefont {S.}~\bibnamefont
  {Clesse}}\ and\ \bibinfo {author} {\bibfnamefont {J.}~\bibnamefont
  {Garc{\'i}a-Bellido}},\ }\href {\doibase 10.1016/j.dark.2016.10.002}
  {\bibfield  {journal} {\bibinfo  {journal} {Phys. Dark Univ.}\ }\textbf
  {\bibinfo {volume} {15}},\ \bibinfo {pages} {142} (\bibinfo {year} {2017})},\
  \Eprint {http://arxiv.org/abs/1603.05234} {arXiv:1603.05234 [astro-ph.CO]}
  \BibitemShut {NoStop}%
\bibitem [{\citenamefont {Bird}\ \emph {et~al.}(2016)\citenamefont {Bird},
  \citenamefont {Cholis}, \citenamefont {Mu{\~n}oz}, \citenamefont
  {Ali-Ha{\"i}moud}, \citenamefont {Kamionkowski}, \citenamefont {Kovetz},
  \citenamefont {Raccanelli},\ and\ \citenamefont {Riess}}]{Bird:2016dcv}%
  \BibitemOpen
  \bibfield  {author} {\bibinfo {author} {\bibfnamefont {S.}~\bibnamefont
  {Bird}}, \bibinfo {author} {\bibfnamefont {I.}~\bibnamefont {Cholis}},
  \bibinfo {author} {\bibfnamefont {J.~B.}\ \bibnamefont {Mu{\~n}oz}}, \bibinfo
  {author} {\bibfnamefont {Y.}~\bibnamefont {Ali-Ha{\"i}moud}}, \bibinfo
  {author} {\bibfnamefont {M.}~\bibnamefont {Kamionkowski}}, \bibinfo {author}
  {\bibfnamefont {E.~D.}\ \bibnamefont {Kovetz}}, \bibinfo {author}
  {\bibfnamefont {A.}~\bibnamefont {Raccanelli}}, \ and\ \bibinfo {author}
  {\bibfnamefont {A.~G.}\ \bibnamefont {Riess}},\ }\href {\doibase
  10.1103/PhysRevLett.116.201301} {\bibfield  {journal} {\bibinfo  {journal}
  {Phys.~Rev.~Lett.}\ }\textbf {\bibinfo {volume} {116}},\ \bibinfo {pages}
  {201301} (\bibinfo {year} {2016})},\ \Eprint
  {http://arxiv.org/abs/1603.00464} {arXiv:1603.00464 [astro-ph.CO]}
  \BibitemShut {NoStop}%
\bibitem [{\citenamefont {Chen}\ \emph {et~al.}(2017)\citenamefont {Chen},
  \citenamefont {Holz}, \citenamefont {Miller}, \citenamefont {Evans},
  \citenamefont {Vitale},\ and\ \citenamefont {Creighton}}]{Chen:2017wpg}%
  \BibitemOpen
  \bibfield  {author} {\bibinfo {author} {\bibfnamefont {H.-Y.}\ \bibnamefont
  {Chen}}, \bibinfo {author} {\bibfnamefont {D.~E.}\ \bibnamefont {Holz}},
  \bibinfo {author} {\bibfnamefont {J.}~\bibnamefont {Miller}}, \bibinfo
  {author} {\bibfnamefont {M.}~\bibnamefont {Evans}}, \bibinfo {author}
  {\bibfnamefont {S.}~\bibnamefont {Vitale}}, \ and\ \bibinfo {author}
  {\bibfnamefont {J.}~\bibnamefont {Creighton}},\ }\href@noop {} {\  (\bibinfo
  {year} {2017})},\ \Eprint {http://arxiv.org/abs/1709.08079} {arXiv:1709.08079
  [astro-ph.CO]} \BibitemShut {NoStop}%
\bibitem [{\citenamefont {Lousto}\ and\ \citenamefont
  {Healy}(2020)}]{Lousto:2020tnb}%
  \BibitemOpen
  \bibfield  {author} {\bibinfo {author} {\bibfnamefont {C.~O.}\ \bibnamefont
  {Lousto}}\ and\ \bibinfo {author} {\bibfnamefont {J.}~\bibnamefont {Healy}},\
  }\href {\doibase 10.1103/PhysRevLett.125.191102} {\bibfield  {journal}
  {\bibinfo  {journal} {Phys. Rev. Lett.}\ }\textbf {\bibinfo {volume} {125}},\
  \bibinfo {pages} {191102} (\bibinfo {year} {2020})},\ \Eprint
  {http://arxiv.org/abs/2006.04818} {arXiv:2006.04818 [gr-qc]} \BibitemShut
  {NoStop}%
\bibitem [{\citenamefont {Abbott}\ \emph {et~al.}(2017)\citenamefont {Abbott}
  \emph {et~al.}}]{TheLIGOScientific:2017qsa}%
  \BibitemOpen
  \bibfield  {author} {\bibinfo {author} {\bibfnamefont {B.}~\bibnamefont
  {Abbott}} \emph {et~al.} (\bibinfo {collaboration} {LIGO Scientific,
  Virgo}),\ }\href {\doibase 10.1103/PhysRevLett.119.161101} {\bibfield
  {journal} {\bibinfo  {journal} {Phys. Rev. Lett.}\ }\textbf {\bibinfo
  {volume} {119}},\ \bibinfo {pages} {161101} (\bibinfo {year} {2017})},\
  \Eprint {http://arxiv.org/abs/1710.05832} {arXiv:1710.05832 [gr-qc]}
  \BibitemShut {NoStop}%
\bibitem [{\citenamefont {Abbott}\ \emph
  {et~al.}(2019{\natexlab{b}})\citenamefont {Abbott} \emph
  {et~al.}}]{Abbott:2019ebz}%
  \BibitemOpen
  \bibfield  {author} {\bibinfo {author} {\bibfnamefont {R.}~\bibnamefont
  {Abbott}} \emph {et~al.} (\bibinfo {collaboration} {LIGO Scientific,
  Virgo}),\ }\href@noop {} {\  (\bibinfo {year} {2019}{\natexlab{b}})},\
  \Eprint {http://arxiv.org/abs/1912.11716} {arXiv:1912.11716 [gr-qc]}
  \BibitemShut {NoStop}%
\bibitem [{\citenamefont {Gow}\ \emph {et~al.}(2020)\citenamefont {Gow},
  \citenamefont {Byrnes}, \citenamefont {Hall},\ and\ \citenamefont
  {Peacock}}]{Gow:2019pok}%
  \BibitemOpen
  \bibfield  {author} {\bibinfo {author} {\bibfnamefont {A.~D.}\ \bibnamefont
  {Gow}}, \bibinfo {author} {\bibfnamefont {C.~T.}\ \bibnamefont {Byrnes}},
  \bibinfo {author} {\bibfnamefont {A.}~\bibnamefont {Hall}}, \ and\ \bibinfo
  {author} {\bibfnamefont {J.~A.}\ \bibnamefont {Peacock}},\ }\href {\doibase
  10.1088/1475-7516/2020/01/031} {\bibfield  {journal} {\bibinfo  {journal}
  {JCAP}\ }\textbf {\bibinfo {volume} {2001}},\ \bibinfo {pages} {031}
  (\bibinfo {year} {2020})},\ \Eprint {http://arxiv.org/abs/1911.12685}
  {arXiv:1911.12685 [astro-ph.CO]} \BibitemShut {NoStop}%
\bibitem [{\citenamefont {Vaskonen}\ and\ \citenamefont
  {Veerm{\"a}e}(2019)}]{Vaskonen:2019jpv}%
  \BibitemOpen
  \bibfield  {author} {\bibinfo {author} {\bibfnamefont {V.}~\bibnamefont
  {Vaskonen}}\ and\ \bibinfo {author} {\bibfnamefont {H.}~\bibnamefont
  {Veerm{\"a}e}},\ }\href@noop {} {\  (\bibinfo {year} {2019})},\ \Eprint
  {http://arxiv.org/abs/1908.09752} {arXiv:1908.09752 [astro-ph.CO]}
  \BibitemShut {NoStop}%
\bibitem [{\citenamefont {Kocsis}\ \emph {et~al.}(2018)\citenamefont {Kocsis},
  \citenamefont {Suyama}, \citenamefont {Tanaka},\ and\ \citenamefont
  {Yokoyama}}]{Kocsis:2017yty}%
  \BibitemOpen
  \bibfield  {author} {\bibinfo {author} {\bibfnamefont {B.}~\bibnamefont
  {Kocsis}}, \bibinfo {author} {\bibfnamefont {T.}~\bibnamefont {Suyama}},
  \bibinfo {author} {\bibfnamefont {T.}~\bibnamefont {Tanaka}}, \ and\ \bibinfo
  {author} {\bibfnamefont {S.}~\bibnamefont {Yokoyama}},\ }\href {\doibase
  10.3847/1538-4357/aaa7f4} {\bibfield  {journal} {\bibinfo  {journal}
  {Astrophys. J.}\ }\textbf {\bibinfo {volume} {854}},\ \bibinfo {pages} {41}
  (\bibinfo {year} {2018})},\ \Eprint {http://arxiv.org/abs/1709.09007}
  {arXiv:1709.09007 [astro-ph.CO]} \BibitemShut {NoStop}%
\bibitem [{\citenamefont {Boehm}\ \emph {et~al.}(2020)\citenamefont {Boehm},
  \citenamefont {Kobakhidze}, \citenamefont {O'Hare}, \citenamefont {Picker},\
  and\ \citenamefont {Sakellariadou}}]{Boehm:2020jwd}%
  \BibitemOpen
  \bibfield  {author} {\bibinfo {author} {\bibfnamefont {C.}~\bibnamefont
  {Boehm}}, \bibinfo {author} {\bibfnamefont {A.}~\bibnamefont {Kobakhidze}},
  \bibinfo {author} {\bibfnamefont {C.~A.}\ \bibnamefont {O'Hare}}, \bibinfo
  {author} {\bibfnamefont {Z.~S.}\ \bibnamefont {Picker}}, \ and\ \bibinfo
  {author} {\bibfnamefont {M.}~\bibnamefont {Sakellariadou}},\ }\href@noop {}
  {\  (\bibinfo {year} {2020})},\ \Eprint {http://arxiv.org/abs/2008.10743}
  {arXiv:2008.10743 [astro-ph.CO]} \BibitemShut {NoStop}%
\end{thebibliography}%

\end{document}